\documentclass[twocolumn]{aastex631}

\usepackage[whole]{bxcjkjatype}
\usepackage{graphicx}

\usepackage{color}
\usepackage{soul}
\sethlcolor{pink}
\usepackage{amsmath, amssymb}
\usepackage[normalem]{ulem} 
\usepackage{multirow}
\usepackage{hyperref}
\usepackage[colorinlistoftodos,backgroundcolor=yellow,textsize=tiny,textwidth=1.25cm]{todonotes}
\usepackage{comment}
\usepackage{bold-extra}

\newcommand{\second}{{\mathrm{\,s}}}

\newcommand{\km}{{\mathrm{\,km}}}

\newcommand{\kms}{{\km\second^{-1}}}

\newcommand{\au}{{\mathrm{\,au}}}

\newcommand{\pc}{{\mathrm{\,pc}}}

\newcommand{\K}{{\mathrm{\,K}}}

\defcitealias{shu1991}{SRLL}
\defcitealias{shang2020}{I}
\defcitealias{shang_PII}{II}
\defcitealias{shang_PIII}{III}

\shorttitle{ALMASOP Outflows}

\begin{document}

\title{ALMA Survey of Orion Planck Galactic Cold Clumps (ALMASOP): \\ Nested Morphological and Kinematic Structures of Outflows Revealed in SiO and CO Emission}

%\begin{CJK*}{UTF8}{bsmi}

\author[0000-0002-1624-6545]{Chun-Fan Liu (劉君帆)}
\affiliation{Institute of Astronomy and Astrophysics, Academia Sinica, Taipei 106216, Taiwan}

\author[0000-0001-8385-9838]{Hsien Shang (尚賢)}
\affiliation{Institute of Astronomy and Astrophysics, Academia Sinica,  Taipei 106216, Taiwan}

\author[0000-0002-6773-459X]{Doug Johnstone}
\affiliation{NRC Herzberg Astronomy and Astrophysics, 5071 West Saanich Rd, Victoria, BC, V9E 2E7, Canada}
\affiliation{Department of Physics and Astronomy, University of Victoria, Victoria, BC, V8P 5C2, Canada}

\author[0000-0002-0517-1197]{Tsung-Han Ai (艾宗瀚)}
\affiliation{Institute of Astronomy and Astrophysics, Academia Sinica,  Taipei 106216, Taiwan}
\affiliation{Department of Physics, National Central University, 300 Zhongda Road, Zhongli, Taoyuan 320317, Taiwan}

\author{Tsz Ming Lee (李子銘)}
\affiliation{Institute of Astronomy and Astrophysics, Academia Sinica, Taipei 106216, Taiwan}
\affiliation{Department of Physics, The Chinese University of Hong Kong, Shatin, New Territory, Hong Kong, People's Republic of China}

\author[0000-0001-5557-5387]{Ruben Krasnopolsky}
\affiliation{Institute of Astronomy and Astrophysics, Academia Sinica, Taipei 106216, Taiwan}

\author[0000-0001-9304-7884]{Naomi Hirano}
\affiliation{Institute of Astronomy and Astrophysics, Academia Sinica, Taipei 106216, Taiwan}

\author[0000-0002-2338-4583]{Somnath Dutta}
\affiliation{Institute of Astronomy and Astrophysics, Academia Sinica, Taipei 106216, Taiwan}

\author[0000-0002-1369-1563]{Shih-Ying Hsu (許世穎)}
\affiliation{Institute of Astronomy and Astrophysics, Academia Sinica, Taipei 106216, Taiwan}

\author[0000-0002-5845-8722]{Jes\'us Alejandro L\'opez-V\'azquez}
\affiliation{Institute of Astronomy and Astrophysics, Academia Sinica, Taipei 106216, Taiwan}

\author[0000-0003-4603-7119]{Sheng-Yuan Liu (呂聖元)}
\affiliation{Institute of Astronomy and Astrophysics, Academia Sinica, Taipei 106216, Taiwan}

\author[0000-0002-5286-2564]{Tie Liu}
\affiliation{Shanghai Astronomical Observatory, Chinese Academy of Sciences, 80 Nandan Road, Shanghai 200030, People's Republic of China}

\author[0000-0002-8149-8546]{Ken'ichi Tatematsu}
\affiliation{Nobeyama Radio Observatory, National Astronomical Observatory of Japan, National Institutes of Natural Sciences, 462-2 Nobeyama, Minamimaki, Minamisaku, Nagano 384-1305, Japan}
\affiliation{Astronomical Science Program, Graduate Institute for Advanced Studies, SOKENDAI, 2-21-1 Osawa, Mitaka, Tokyo 181-8588, Japan}

\author[0000-0003-2384-6589]{Qizhou Zhang}
\affiliation{Center for Astrophysics, Harvard \& Smithsonian, 60 Garden Street, Cambridge, MA 02138, USA}

\author[0000-0002-6529-202X]{Mark G. Rawlings}
\affiliation{Gemini Observatory/NSF's NOIRLab, 670 N. A'ohoku Place, Hilo, Hawai'i, 96720, USA}

\author[0000-0002-5881-3229]{David Eden}
\affiliation{Armagh Observatory and Planetarium, College Hill, Armagh BT61 9DG, UK}

\author[0000-0003-4659-1742]{Zhiyuan Ren}
\affiliation{National Astronomical Observatories, Chinese Academy of Sciences, Beijing 100101, People’s Republic of China}

\author[0000-0002-7125-7685]{Patricio Sanhueza}
\affiliation{National Astronomical Observatory of Japan, National Institutes of Natural Sciences, 2-21-1 Osawa, Mitaka, Tokyo 181-8588, Japan}
\affiliation{Astronomical Science Program, The Graduate University for Advanced Studies, SOKENDAI, 2-21-1 Osawa, Mitaka, Tokyo 181-8588, Japan}

\author[0000-0003-4022-4132]{Woojin Kwon}
\affiliation{Department of Earth Science Education, Seoul National University, 1 Gwanak-ro, Gwanak-gu, Seoul 08826, Republic of Korea}
\affiliation{SNU Astronomy Research Center, Seoul National University, 1 Gwanak-ro, Gwanak-gu, Seoul 08826, Republic of Korea}

\author[0000-0002-3179-6334]{Chang Won Lee}
\affiliation{Korea Astronomy and Space Science Institute (KASI), 776 Daedeokdae-ro, Yuseong-gu, Daejeon 34055, Republic of Korea}
\affiliation{University of Science and Technology, Korea (UST), 217 Gajeong-ro, Yuseong-gu, Daejeon 34113, Republic of Korea}

\author[0000-0002-4336-0730]{Yi-Jehng Kuan (管一政)}
\affiliation{Institute of Astronomy and Astrophysics, Academia Sinica, Taipei 106216, Taiwan}
\affiliation{Department of Earth Sciences, National Taiwan Normal University, Taipei 116059, Taiwan, R.O.C.}

\author[0000-0002-4086-7632]{Somdeb Bandopadhyay}
\affiliation{Institute of Astronomy and Astrophysics, Academia Sinica, Taipei 106216, Taiwan}

\author[0000-0002-8782-4664]{Miikka S. V\"ais\"al\"a}
\affiliation{Institute of Astronomy and Astrophysics, Academia Sinica, Taipei 106216, Taiwan}

\author[0000-0003-4603-7119]{Chin-Fei Lee (李景輝)}
\affiliation{Institute of Astronomy and Astrophysics, Academia Sinica, Taipei 106216, Taiwan}

\author[0000-0002-7424-4193]{Indrani Das}
\affiliation{Institute of Astronomy and Astrophysics, Academia Sinica, Taipei 106216, Taiwan}

%\end{CJK*}

\correspondingauthor{Chun-Fan Liu, Hsien Shang}
\email{cfliu,shang@asiaa.sinica.edu.tw}

\begin{abstract}

The Atacama Large Millimeter/submillimeter Array Survey of Orion Planck Galactic Cold Clumps (ALMASOP) reveals complex nested morphological and kinematic features of molecular outflows through the CO ($J=2-1$) and SiO ($J=5-4$) emission. We characterize the jet and outflow kinematics of the ALMASOP sample in four representative sources (HOPS 10, 315, 358, and G203.21-11.20W2) through channel maps and position--velocity diagrams (PVDs) parallel and transverse to the outflow axes. The combined CO and SiO emission exhibits the coexistence of the conventional extremely--high-velocity (EHV) jets and shell-like low-velocity (LV) cavity walls and new features. More complex, nested bubble-like and filamentary structures in the images and channel maps, triangle-shaped regions near the base of the parallel PVDs, and regions composed of rhombus/oval shapes in the transverse PVDs, are also evident. Such features find natural explanations within the bubble structure of the unified model of jet, wind, and ambient medium. The reverse shock cavity is revealed on the PVD base regions, and other features naturally arise within the dynamic postshock region of magnetic interaction. The finer nested shells observed within the compressed wind region reveal previously unnoticed shocked emission between the jet and the conventional large cavity walls.  These pseudopulse-produced filamentary features connect to the jet-like knotty blobs, creating an impression of episodicity in mass ejection. SiO emission is enhanced downstream of the reverse shock boundary, with jet-like excitation conditions. Combined, these observed features reveal the extended structures induced by the magnetic interplay between a jet-bearing magnetized wide-angle wind and its ambient magnetized surrounding medium.

\end{abstract}

\section{Introduction} \label{sec:intro}

Molecular outflows from young stellar objects (YSOs) are probes of the stellar vicinity and its immediate surrounding environment through the interaction between the jets, winds, and ambient medium. Observed around Class 0 to early Class II YSOs, molecular outflows exist across the evolutionary stages of star formation with various morphological and kinematic properties \citep[e.g.,][]{arce2006}.

Interferometric observations have been used to reveal the morphological and kinematic structures within molecular outflows. 
Molecular outflows observed in CO emission are often dissected into roughly two morphologically distinct components when viewed through channel maps: a shell-like wide-angle outflow lobe in lower velocity channels surrounding jet-like emission in the highest velocity channels.
Such a conventional jet--shell structure has been revealed in several Class 0/I outflow sources by the Plateau de Bure Interferometer (PdBI) and the Submillimeter Array (SMA), including HH 211 \citep{gueth1999,palau2006}, IRAS 04166+2706 \citep{santiago-garcia2009,wang2014}, and L1448C-N \citep{hirano2010}.
The jet-like structure is often referred to as the extremely--high-velocity (EHV) component, and the outer shell structure is referred to as the low-velocity (LV) component, which, along with the intermediate-velocity (IV) wing, is part of the standard--high-velocity (SHV) component of the molecular outflow \citep{bachiller1996,bally2016,tafalla2010}.
Spectroscopically, the LV component has a profile peaked close to the systemic velocity with the IV wing toward higher velocities, whereas the EHV component usually demonstrates a distinct peak.
The EHV jet typically exhibits higher density than the IV wing in that the SiO transitions tend to be present in the EHV jet but absent in the IV wing \citep{tafalla2010}. 

Detailed analyses of jet and outflow signatures require careful consideration of features produced by the EHV jet, the LV outflow cavities, and the intermediate regions connecting the two.
With higher sensitivity and resolution, the Atacama Large Millimeter/submillimeter Array (ALMA) has revealed more complex kinematic features in the outflow lobes. These structures include multiple molecular layered features shown in Class 0/I sources \citep[e.g.,][]{plunkett2015b,tafalla2017MmSAI,lee2018_hh212} and low-velocity molecular outflows from Class I/II sources with known optical jets \citep[e.g.,][]{louvet2018,zhang2019,devalon2020,guedel2018}. Such complex kinematics have provoked subsequent investigations and examinations as to how they can be consistently interpreted.

From a theoretical perspective, \citet{shu1991} extended our understanding of the nature of these molecular outflows by predicting the morphology and kinematics of outflows driven by a spherically distributed radial wind with a fundamental momentum-conserving law, yielding a momentum-conserving thin shell with a ``Hubble-law'' velocity pattern. Alternatively, a jet-driven mechanism was proposed to account for outflows with large length-to-width ratios. \citet{lee2001} found reasonable results by fitting these kinematic properties of a hydrodynamic thin shell driven by both a wind and a pure cylindrical jet against radio observations.

The unified model of bipolar outflows developed by \cite{shang2006} naturally produces the features expected in the jet-driven and the wind-driven momentum-conserving shells. In the new framework of the magnetic interplay between magnetized winds and ambient toroids advanced by \citet{shang2020}, combinations within the parameter space of the wind magnetization (in terms of Alfv\'enic Mach number of the wind $M_\mathrm{A}$), ambient toroid flattening $n$, and ambient magnetization produce systematic nested kinematic and morphological features, for which \citet{shang_PII} advanced a wide variety of kinematic features potentially observable within molecular outflows resulting from the interplay. 
Position--velocity diagrams (PVDs) parallel and transverse to the outflow axis are determined to be especially powerful in probing the key kinematic features predicted in \citet{shang_PII}, and for probing mixing of the entrained ambient material with the pristine wind in the shocked region \citep{shang2020}.

\citet{ai2024} demonstrated how the multiple velocity components of HH 30 observed at both optical and radio wavelengths can be unified through an inner atomic wind interacting with the ambient surroundings without the requirement to invoke a separate slow molecular wind.
The ALMA $^{12}$CO channel maps of the HH 30 outflow system show multiple filamentary structures surrounding the EHV optical jet. The observations also reveal nested kinematics of layered velocity components as intersecting ovals on the transverse PV diagrams, predicted in \citet{shang_PII}.
The low-velocity $\sim5$\,$\kms$ CO emission is therefore taken to be the ambient material mixed into the outflow lobe by the high-velocity $\sim100$\,$\kms$ wide-angle magnetized atomic wind, with the velocity differences mainly due to the density contrast between the wind and surroundings together entrained via momentum-conserving shell formation.

This work utilizes channel maps and position--velocity diagrams to investigate the morphological and kinematic properties of a selected sample of outflows from the Atacama Large Millimeter/submillimeter Array Survey of Orion Planck Galactic Cold Clumps \citep[ALMASOP,][]{dutta2020}. ALMASOP covers the Orion A/B and $\lambda$-Orionis regions and provides an unbiased survey of Class 0 and I sources, all at a similar distance of $\sim400\pc$. An overview of the survey, including the properties of 1.3 mm continua and detection and identification of outflows through $^{12}$CO and SiO, has been summarized by \citet{dutta2020}. Subsequent analyses of the dataset have been performed on aspects including chemistry, hot corinos, and multiplicity \citep[e.g.,][]{hsu2022,luo2022,sahu2023}. A series of outflow studies have also been undertaken for this sample, including detection and analysis of SiO jets from likely the earliest Class 0 phase and the late Class I phase \citep{dutta2022,dutta2022_G208W} and analysis of 6 SiO jets with distinct knot structure along the jet allowing the derivation of inclination angles, jet velocities, and potential ejection periods \citep{jhan2022, dutta2024}.

The outline of this work is as follows. Section \ref{sec:obs} summarizes the properties of the ALMASOP dataset and the analyses performed. Section \ref{sec:results} demonstrates the properties of four representative outflow sources in the ALMASOP sample in both CO and SiO emission by identifying the morphological and kinematic features in the outflow lobes. Section \ref{sec:discussion} discusses the implication and interpretation of the outflow kinematic structures. A historical account of the theory and simulation aspects of molecular outflows is provided in Section \ref{subsec:history}. The advancement of the unified wind-blown bubble model and its applications to the observed features are discussed in Section \ref{subsec:unified_wind_model}, starting with a recapitulation of the selected samples in \ref{subsubsec:unified_wind_applications}. Its insight to the previously unnoticed signature of the reverse-shock cavity and to the nested kinematic signatures are discussed in Sections \ref{subsec:RSCavity} and \ref{subsec:pv_discussion}. Section \ref{subsec:sio_formation} discusses the origins of the SiO emission in embedded young stellar systems. Section \ref{subsec:episodicity} discusses the appearance of apparent and real episodicity in molecular outflows using this subset of the ALMASOP sample as examples. Next, Section \ref{subsec:conundrums_conventional} discusses the problems of applying the conventional pictures of thin-shell, jet-driven, and wind-driven outflow models, and discusses the difficulties of the alternative slow wind model. Section \ref{subsec:ensemble} concludes the discussion with prospective to the broader ensemble of ALMASOP outflows. Section \ref{sec:summary} summarizes our findings.

\section{Observational Analysis} \label{sec:obs}

This work investigates the kinematic features of molecular outflows detected in the ALMASOP sample. 

\subsection{Observational Datasets} \label{subsec:obsdata}

The ALMASOP observations were performed with the ALMA Band 6 setup from 2018 October through 2019 January (Project ID: 2018.1.00302.S). Details of the observational setup and detected line transitions have been summarized by \citet{dutta2020}. We extract the molecular transitions of $^{12}$CO ($J=2-1$) and SiO ($J=5-4$) for the kinematic studies of the ALMASOP outflow sources, and of C$^{18}$O ($J=2-1$), N$_2$D$^+$ ($J=3-2$), and H$_2$CO ($J_{{K_A},{K_C}}=3_{0,3}-2_{0,2}$) for investigations of circumstellar structures. C$^{18}$O is also used for determination of the systemic velocities \citep[e.g.,][]{dutta2024}. Datasets were obtained using three ALMA observing configurations, i.e., two for the 12-m (TM1 and TM2) and one for the 7-m Atacama Compact Array (ACA). 
The datacubes used in this paper combine all three configurations, except for N$_2$D$^+$  where the combined datacubes filtered out the large-scale structure in the spatial distribution and therefore the ACA-only datasets are analysed. The velocity sampling of the datacubes was rebinned from $\sim1.5\kms$ to $\sim2\kms$ to increase the signal-to-noise ratios \citep{dutta2020}. Transitions and beam sizes of the lines reported in this work are summarized in Table \ref{tab:obs_mollines}.

Out of 72 ALMASOP objects, a total of 31 sources drive well-defined CO outflows, 19 of which show SiO emission in their outflows. These outflows emanate from mostly Class 0 sources, but also including four Class I sources, based on $T_{\rm bol}$ classification \citep{dutta2020}. The sample spans a variety of $T_{\rm bol}$ and $L_{\rm bol}$, providing a range of various morphological and kinematic features.

\begin{deluxetable*}{cccccc}
\tabletypesize{\scriptsize}
\tablewidth{0pt} 
\tablecaption{Analyzed Molecular Transitions \label{tab:obs_mollines}}
\tablehead{
\colhead{Species} & \colhead{Transition}& \colhead{Rest Frequency} & \colhead{Data Combination} & \colhead{Beam Size} & \colhead{Pixel Scale}
} 
\startdata
SiO         & $J=5-4$ & 217.105 GHz & TM1+TM2+ACA & $0\farcs42\times0\farcs38$ & 0\farcs06 \\
H$_2$CO     & $J_{{K_A},{K_C}}=3_{0,3}-2_{0,2}$ & 218.222 GHz & TM1+TM2+ACA & $0\farcs43\times0\farcs36$ & 0\farcs06 \\
C$^{18}$O   & $J=2-1$ & 219.560 GHz & TM1+TM2+ACA & $0\farcs46\times0\farcs39$ & 0\farcs06 \\
$^{12}$CO   & $J=2-1$ & 230.538 GHz & TM1+TM2+ACA & $0\farcs41\times0\farcs35$ & 0\farcs06 \\
N$_2$D$^+$  & $J=3-2$ & 231.322 GHz & ACA & $7\farcs35\times4\farcs20$ & 0\farcs78 \\ 
\enddata
\end{deluxetable*}

\subsection{Kinematic Analysis} \label{subsec:obsanalysis}

We identified the driving sources and position angles of the ALMASOP outflows. The datacubes were first integrated in the frequency domain to produce Moment 0 maps. The outflow position angles of the individual sources were obtained by visually inspecting the $^{12}$CO and SiO Moment 0 maps using an OpenCV-utilized in-house Python script described by \citet{ai2024}.
The extensions of the outflow lobes along and across the outflow axis were also visually inspected and defined through the $^{12}$CO Moment 0 maps using the same script. The source positions were rectified using 1.3 mm continuum results as compiled in Table 5 of \citet{dutta2020}.

The $^{12}$CO and SiO datacubes were analyzed along each of their three dimensions. The $^{12}$CO and SiO datacubes were first rotated using the measured individual position angles such that the outflow axes are aligned along the vertical direction with the redshifted lobe pointing upward. Datacubes were binned over the frequency axis to produce integrated channel maps in order to examine the spatial morphological structures among various velocity components. Position--velocity diagrams (PVDs) were obtained both parallel to the outflow axis (parallel PVDs) and perpendicular to the outflow axis (transverse PVDs) by averaging across and along the outflow axis, respectively. For the parallel PVDs, one set of ``narrow'' coverage across $\sim0\farcs6$ was used for investigating the central CO and SiO jet zones, and another set of ``broad'' coverage covering the entire
CO outflow lobes was used for investigating the full CO kinematics. For the transverse PVDs, those of the blueshifted and redshifted lobes were integrated to investigate the overall kinematic structure of the lobes. The systemic velocities of individual sources were determined by identifying the peak frequencies of the integrated spectra of C$^{18}$O around each source \citep[cf.][]{jhan2022,dutta2024}. 

We interrogate the $^{12}$CO and SiO datacubes in the manner of recent explorations of the Class I/II HH\,30 jet--outflow system {by \citet{ai2024}, where HST optical data and ALMA radio data were combined and analysed. 
Their work involves optical forbidden line profile analysis and identification of features shown in the molecular line emission datacubes through channel maps and parallel as well as transverse PVDs.
They showed that HH\,30 exhibits nested $\sim5\kms$ LV $^{12}$CO filamentary structures encompassing the bright optical EHV jet with a large full line width of $\sim200\kms$, and the whole outflow lobe is surrounded by very--low-velocity (VLV) $\sim2\kms$ molecular magnetized ambient material also traced by $^{12}$CO\@. The nested structure of HH 30 manifests a magnetized wind bubble formed by an atomic wide-angle wind interacting with a magnetized molecular ambient environment as depicted by \citet{shang2020,shang_PII}. With such interesting and varied kinematic structure in mind, we investigate the morphological and kinematic features revealed in the ALMASOP outflow sample. 

\subsection{Sources Selected in this Work} \label{subsec:oursources}

Starting with the ALMASOP outflow sample, we surveyed the 19 systems with CO and SiO outflows and searched for sources with distinct features of the nested jet--outflow structures shown in their channel maps and PVDs.
Out of the 19 outflow sources, 4 show the most clear nested kinematic structures among a range of various inclinations and thus were selected to demonstrate our kinematic analysis and interpretation. Three of these sources are included in the Herschel HOPS catalogue \citep{furlan2016}, HOPS 10, 315, and 358, while the final source retains its ALMASOP designation G203.21-11.20W2 (or, in short, G203W2).
Left for future works are sources where the outflow axes are almost in the plane of the sky (kinematic structure not well resolved in the current setting), sources with spatially unresolved SiO emission, and those with large (length-to-width ratio $\gtrsim10$) and compact (extending $\lesssim5\arcsec$) CO/SiO emission. 

Our four analysed outflow sources were also reported by \citet{dutta2024} utilizing an independent investigation on the possible ejection episodicities in the jets and outflows.
Table \ref{tab:outflows_src_prop} summarizes the properties of the four outflow sources selected from the ALMASOP sample, adopted from \citet{dutta2020}, including the continuum peak positions, infrared spectral classes, and associated HOPS numbers. 
The position angles of the outflow axes, obtained by $^{12}$CO morphology, and the LSR systemic velocities, obtained by C$^{18}$O spectra, are also provided in Table \ref{tab:outflows_src_prop}. Inclination angles of the outflows, as defined by the angle between the line of sight and the outflow axis, are adapted from the recent compilations of \citet{jhan2022} and \citet{dutta2024}.

\begin{deluxetable*}{lrccrccc}
\tabletypesize{\scriptsize}
\tablewidth{0pt} 
\tablecaption{Properties of the Outflow Source Samples \label{tab:outflows_src_prop}}
\tablehead{
\colhead{Source Name} & \colhead{HOPS ID} & \colhead{RA (J2000)}& \colhead{Dec (J2000)} & \colhead{PA$_\mathrm{red}$ ($^\circ$)} & \colhead{$i$} ($^\circ$) & \colhead{$v_{\rm sys}$ ($\kms$)} & \colhead{Protostellar Class} 
} 
\colnumbers
\startdata 
{G205.46-14.56S3   } & 315 & 05h46m03.63s & -00d14m49.57s & $227.3$ & $40^\circ$ & $10\pm2$ & I \\
{G209.55-19.68S2   } &  10 & 05h35m09.05s & -05d58m26.87s & $ 33.6$ & $62^\circ$ & $ 8\pm2$ & 0 \\
{G203.21-11.20W2   } & \nodata & 05h53m39.51s & +03d22m23.85s & $149.1$ & $72^\circ$ & $10\pm2$ & 0 \\
{G205.46-14.56S1\_A} & 358 & 05h46m07.26s & -00d13m30.23s & $163.4$ & $89^\circ$ & $10\pm2$ & 0 \\
\enddata
\tablecomments{(1) Source name, (2) HOPS identification number \citep{furlan2016}, (3) Continuum peak RA, (4) Continuum peak Dec, (5) Position angle of the redshifted lobe, (6) Inclination angle defined by the angle between the line of sight and outflow axis, adopted from \citet{jhan2022} and \citet{dutta2024}, (7) Systemic velocity with respect to the LSR, obtained from C$^{18}$O spectrum, (8) Infrared protostellar class. Source names, their RA and Dec values, and protostellar classes are following \citet{dutta2020}.}
\end{deluxetable*}

\section{Features Observed in the Representative Outflow Sources} \label{sec:results}

The representative sources demonstrated here are G205.46-14.56S3 (HOPS 315), G209.55-19.68S2 (HOPS 10), G203.21-11.20W2, and G205.46-14.56S1 (HOPS 358), in order of increasing inclination angle \citep[values adopted from][]{jhan2022} toward the edge-on orientation. These sources will be hereafter referred to by their HOPS identifiers, or G203W2 for G203.21-11.20W2, which does not have a HOPS designation.

Figure \ref{fig:jet-cavity_maps} showcases the CO integrated maps of these 4 sources. The SiO emission, separated into the integrated blueshifted and redshifted components, is overlaid on the CO integrated maps as blue and red contours, respectively. CO and SiO emission can be seen clearly to trace the outflows associated with these systems. Integrated intensities over $\pm5\kms$ from the systemic velocities for N$_2$D$^+$, H$_2$CO, and C$^{18}$O are overlaid along with CO emission as contours in the subsequent panels. 

\begin{figure*}
\plotone{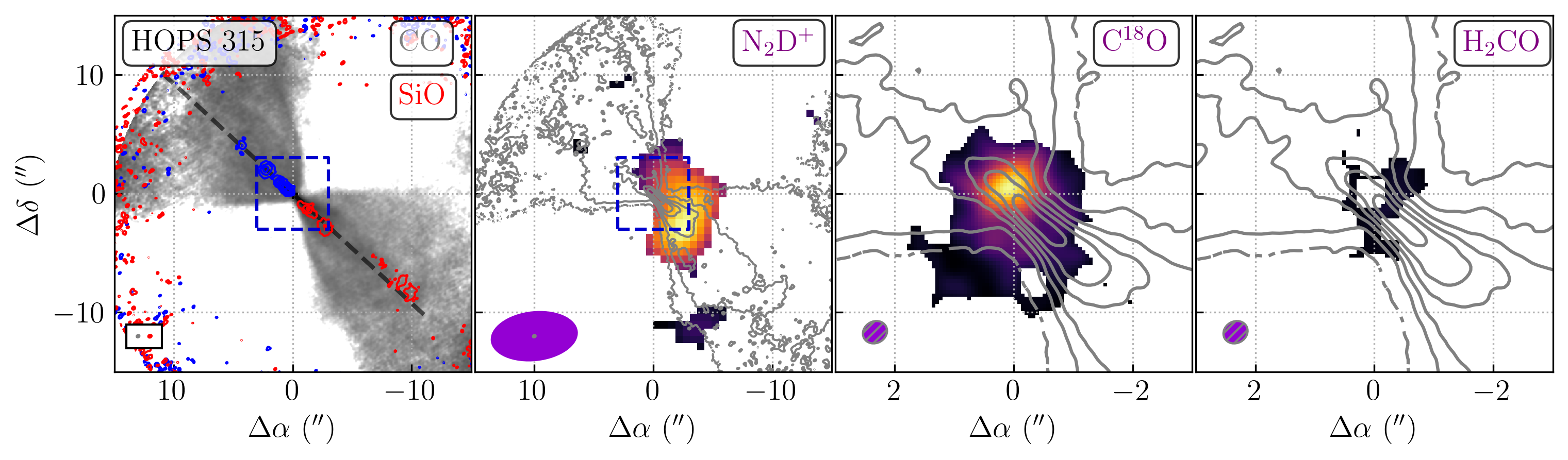}\\
\plotone{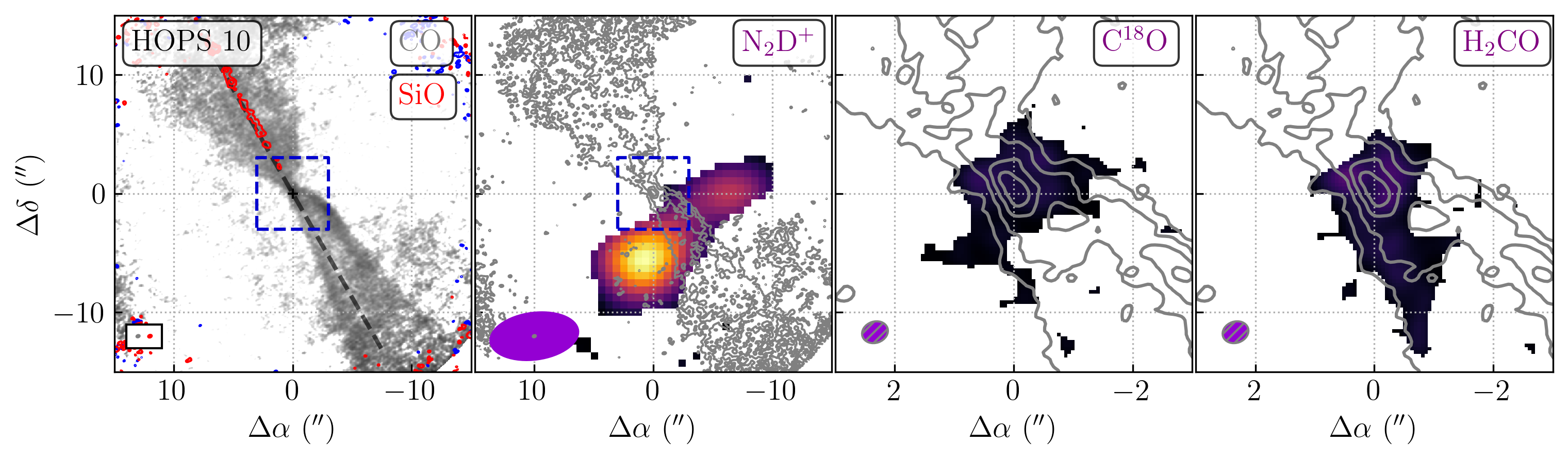}\\
\plotone{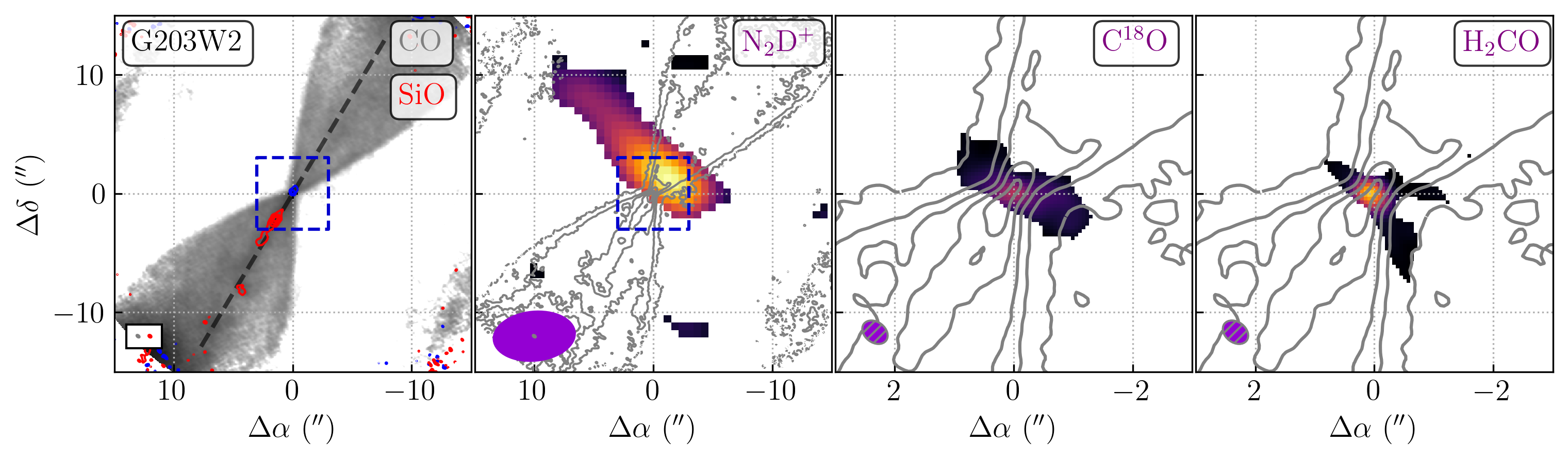}\\
\plotone{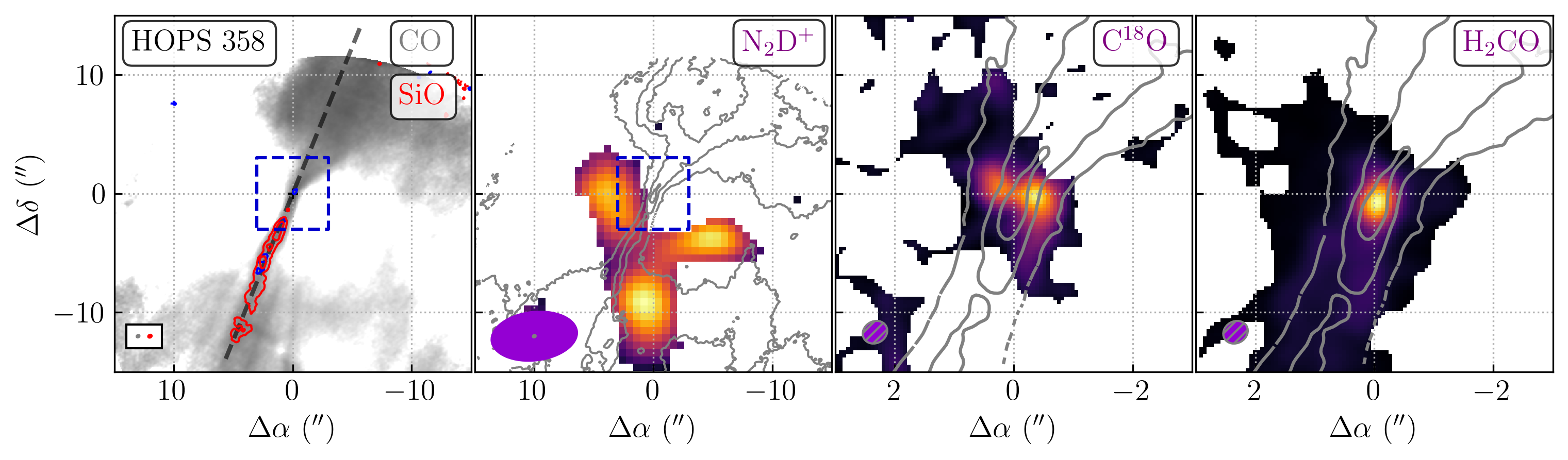}
\caption{Integrated maps of various emission lines of the selected sources G205.46-14.56S3 (HOPS 315), G209.55-19.68S2 (HOPS 10), G203.21-11.20W2 (G203W2), and G205.46-14.56S1 (HOPS 358), from top to bottom rows, in an increasing order of inclination angle, i.e., going toward the edge-on disk orientation. For each of the sources, the leftmost panel shows $^{12}$CO in the background grey-scale colormap, and SiO blueshifted and redshifted lobes in blue and red contours, with synthesized beam patterns for $^{12}$CO and SiO in grey and red ellipses in the lower left corner. For the next three panels, the $^{12}$CO is shown as dark grey contours, and the colormaps show N$_2$D$^+$, C$^{18}$O, and H$_2$CO, respectively, in which the $^{12}$CO beam patterns are shown as grey hatched ellipses on top of the beam patterns (shown in purple ellipses) of the other three lines. Gaussian smoothing has been applied on the N$_2$D$^+$, C$^{18}$O, and H$_2$CO maps with a Gaussian standard deviation $\sigma_G$ of 1 pixel. The two leftmost panels have fields of view of $30\arcsec\times30\arcsec$ with dashed boxes indicating the $6\arcsec\times6\arcsec$ fields of view for the two rightmost panels. All the panels are centered at the continuum peaks of the individual sources, indicated by the cross sign. The thick black lines in the leftmost $^{12}$CO colormaps are the position angles on which the parallel and transverse PVDs in Figures \ref{HOPS315_PVDs}, \ref{HOPS10_PVDs}, \ref{G203W2_PVDs}, and \ref{HOPS358_PVDs} are based.}\label{fig:jet-cavity_maps}
\end{figure*}

\subsection{Circumstellar Structure} \label{subsec:circumstellar_structure}

The local circumstellar structures observed around the outflow sources are mainly traced by N$_2$D$^+$, C$^{18}$O, and H$_2$CO transitions. These line transitions are confined within $\pm5\kms$ from the systemic velocities for most sources. 

N$_2$D$^+$ generally traces cold environments within the circumstellar ambient region \citep[see, e.g.,][]{tychoniec2021}.
As shown in Figure \ref{fig:jet-cavity_maps}, the N$_2$D$^+$ structures in HOPS 10 and G203W2 appear relatively rectangular, whereas for HOP 315 it appears to be closer to triangular with the apices pointing towards the outflow driving source. 
The effect of viewing angle may produce this change in appearance, as HOPS 315 has a smaller inclination than the former two cases (Table \ref{tab:outflows_src_prop}). The almost edge-on system HOPS 358 appears to have an X-shaped N$_2$D$^+$ structure that might confine the large-scale CO opening. Similar flattened N$_2$D$^+$ structure has also been found across another edge-on system, G208.89–20.04Walma \citep{dutta2022_G208W}.

Also seen in Figure \ref{fig:jet-cavity_maps}, C$^{18}$O generally traces the compact and dense part of the circumstellar envelope. It may also trace dense ambient regions just outside of the outflow lobe, delineating the outer surface of the outflow cavity. In G203W2, C$^{18}$O mainly traces the compact flattened structure surrounding the driving source, whereas in HOPS 358, HOPS 315, and HOPS 10, more extended ambient material, encompassing the $^{12}$CO outflow lobes, is traced. The dense ambient environment likely shapes the outflow lobes by confining the wind driven from the source.

H$_2$CO generally traces compact envelopes surrounding the base structure of the outflow cavity walls, similar to the appearance of C$^{18}$O\@. It can occasionally trace regions along the outflow lobe, as shown in the almost edge-on systems HOPS 358 and G208.89–20.04Walma \citep{dutta2022_G208W}, and in the high-mass regime as well \citep[e.g.,][]{izumi2023}. In the low-mass cases, H$_2$CO emission is found in the lobe where also SiO emission is detected, revealing a relatively narrow jet-like appearance. H$_2$CO can therefore trace a range of circumstellar components, including the EHV molecular jet \cite[see, e.g.,][]{tychoniec2021}. 

\subsection{Outflow Properties by Source} \label{subsec:selected_samples}

\subsubsection{G205.46-14.56S3 (HOPS 315)}\label{subsubsec:HOPS315}

HOPS 315 is also known as HH 26 IRS, which drives the chain of Herbig--Haro objects HH 26A/C seen in H$_2$ 2.12$\mu$m at distances of $20\arcsec$ and $50\arcsec$ along the position angle of $\sim45^\circ$ \citep{chrysostomou2000}. A small-scale (within $5\arcsec$) monopolar H$_2$ jet along the same position angle was detected by \citet{davis2002}. [\ion{Fe}{2}] 1.644$\mu$m emission was detected toward HH 26A \citep{caratti-o-garatti2006} but not found at the driving source HH 26 IRS \citep{antoniucci2008}; the same situation also applies to [\ion{O}{1}] 63$\mu$m emission \citep{sperling2020}. Large-scale $^{12}$CO ($2-1$) emission was reported by \citet{dunham2014b}.

The outflow of HOPS 315 (G205.46-14.56S3) observed by the ALMASOP survey has been reported by \citet{dutta2022} and \citet{jhan2022}. The driving source was classified as Class I based on its high bolometric temperature \citep[$T_{\rm bol}=180\pm33\K$;][]{furlan2016} although a pair of prominent SiO jets can still be seen, making it possibly a rare occurrence of a powerful SiO jet being driven at a rather evolved stage. The jet has been identified to have at least four knots on both the blueshifted and redshifted sides \citep{dutta2022}. 

Figure \ref{HOPS315_ChMap} shows the CO channel maps of HOPS 315 in color-scale background with SiO channel maps overlaid on top as white contours. Knotty structure can be seen most prominently on the fastest channel maps in CO and SiO, labeled by short horizontal bars. The knots may be identified on the blueshifted side as B1 to B4 and on the redshifted side as R1 to R4. Arcs associated with the positions of the knotty structures, and seen in other velocity channels, are delineated by dotted curves.
From the channel maps, the CO emission shows a large opening angle at close to its systemic velocity of $\sim10\kms$. At velocities away from the systemic velocity, the CO emission becomes conic toward the source and bubble-like downstream. The opening angles for the conic and bubble-like CO structures decrease as the velocities increase. In the parallel PVDs, the CO emission can be seen to be present in the low-velocity shells, the high-velocity jets, and the intermediate region connecting these two components. Combining the CO and SiO on the parallel PVDs, the CO and SiO EHV jet emission can be seen to be connected to the source origin by CO filamentary structures (traced by light-green dotted lines in Figure \ref{HOPS315_PVDs}), most prominently seen for knots R2 to R4. 

More kinematic features are revealed through the parallel PVDs (Figure \ref{HOPS315_PVDs}). On the redshifted side, a large line width at the base of the jet is indicated by the first SiO knot (R1). Such a large line width is also found for the first SiO knot of the blueshifted side (B1). However, SiO was not detected directly toward the base of the jet as conventionally expected but appears to delineate a triangular-shaped SiO cavity in the parallel PV space, traced by yellow dashed lines shown in Figure \ref{HOPS315_PVDs}. 
Neither the SiO or CO emission  appear in the jet portion until downstream of the SiO cavity. These SiO knots, shown beyond the SiO cavity, each appears to show slight velocity slope, somewhat resembling the CO and SiO patterns found in L1448N-C \citep{hirano2010,toledano-juarez2023}. However, the velocity centroids of these knots have a relatively stable pattern showing projected velocity shifts within $\sim20\kms$.

The CO transverse PVD (Figure \ref{HOPS315_PVDs}, lower left) shows emission with the shape of a concave rhombus (gray dotted line), roughly centered on the systemic velocity and source position. Two characteristic ellipses or ovals can be traced within this rhombus (one high-velocity in light-blue dashes, and one low-velocity in light-green dashes). An additional bright oval (black dots) corresponds to even smaller line-of-sight velocities close to the systemic velocity, and extends to larger positions.
The SiO emission appears as two bright knots near the $\pm v$ ends of major axis of the high-velocity oval. The redshifted side has a brighter SiO knot at $\sim75\kms$ with respect to $v_\mathrm{sys}$, and the blueshifted side has a knot at $\sim-80\kms$ with respect to $v_\mathrm{sys}$, tracing the inner axial density portion of HOPS 315.

\begin{figure*}
 \begin{center}
 \includegraphics[angle=-90,scale=0.7]{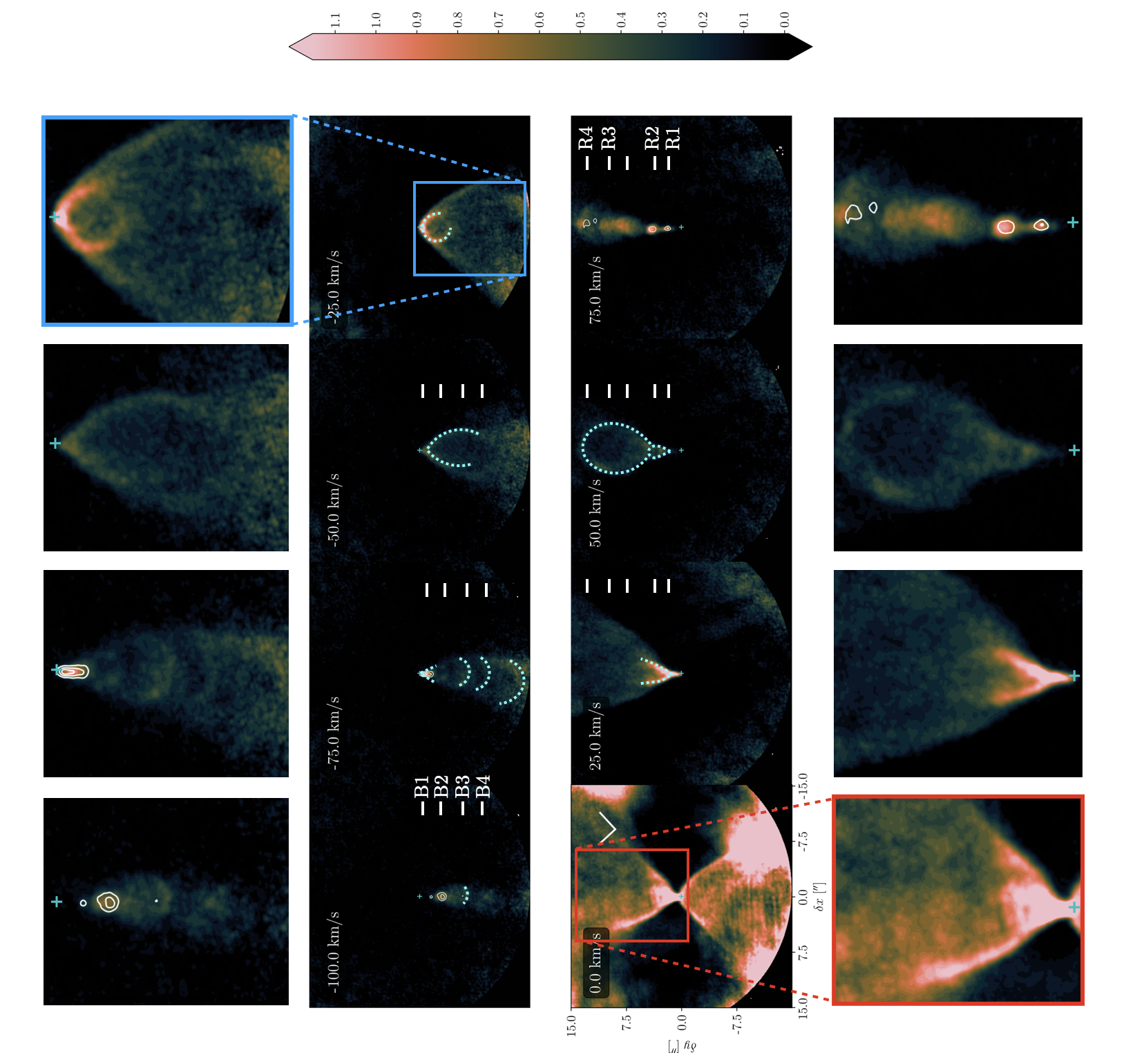}
 \caption{G205.46-14.56S3 (HOPS 315): integrated channel maps with $\sim25\kms$ width per channel shown in the central two rows. The central velocity of each channel is shown in $v_\mathrm{LSR}$. The maps have been rotated so that the redshifted outflow axis is upwards with north--east compass bars in the lower-left panel. The maps show $^{12}$CO in color maps and SiO in white contours in units of Jy\,beam$^{-1}\kms$. Positions of high-velocity SiO blobs identified in \citet{dutta2022,jhan2022} are indicated by short horizontal bars, and arc-like and bubble-like CO structures are delineated by dotted lines. The uppermost and lowermost rows show zoomed-in views of the blueshifted and redshifted lobes with field of views indicated by the blue and red rectangles, respectively. In each of the panels, the nominal position of the star is drawn as a green cross.} 
 \label{HOPS315_ChMap}
 \end{center}
\end{figure*}

\begin{figure*}
 \plottwo{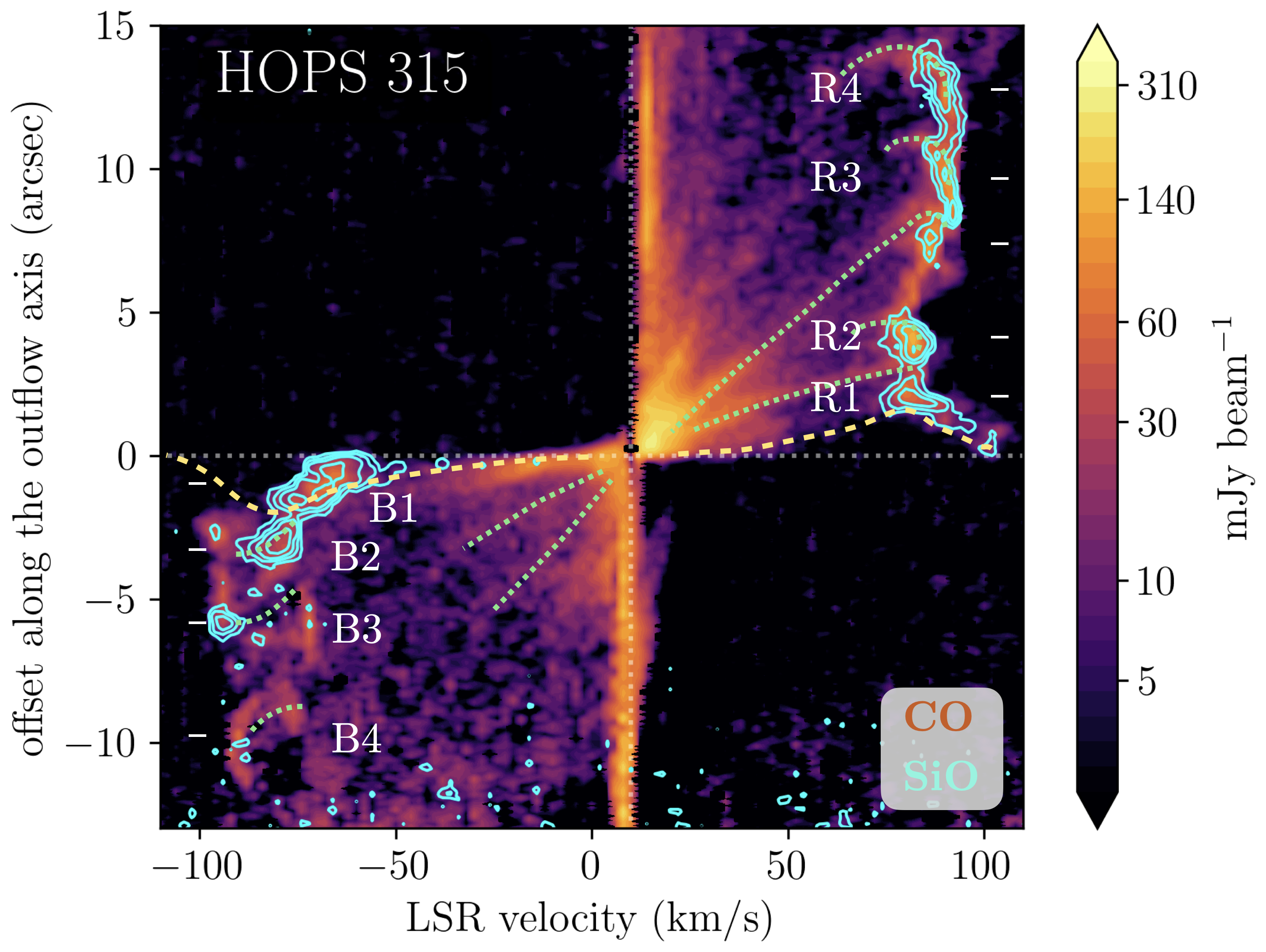}{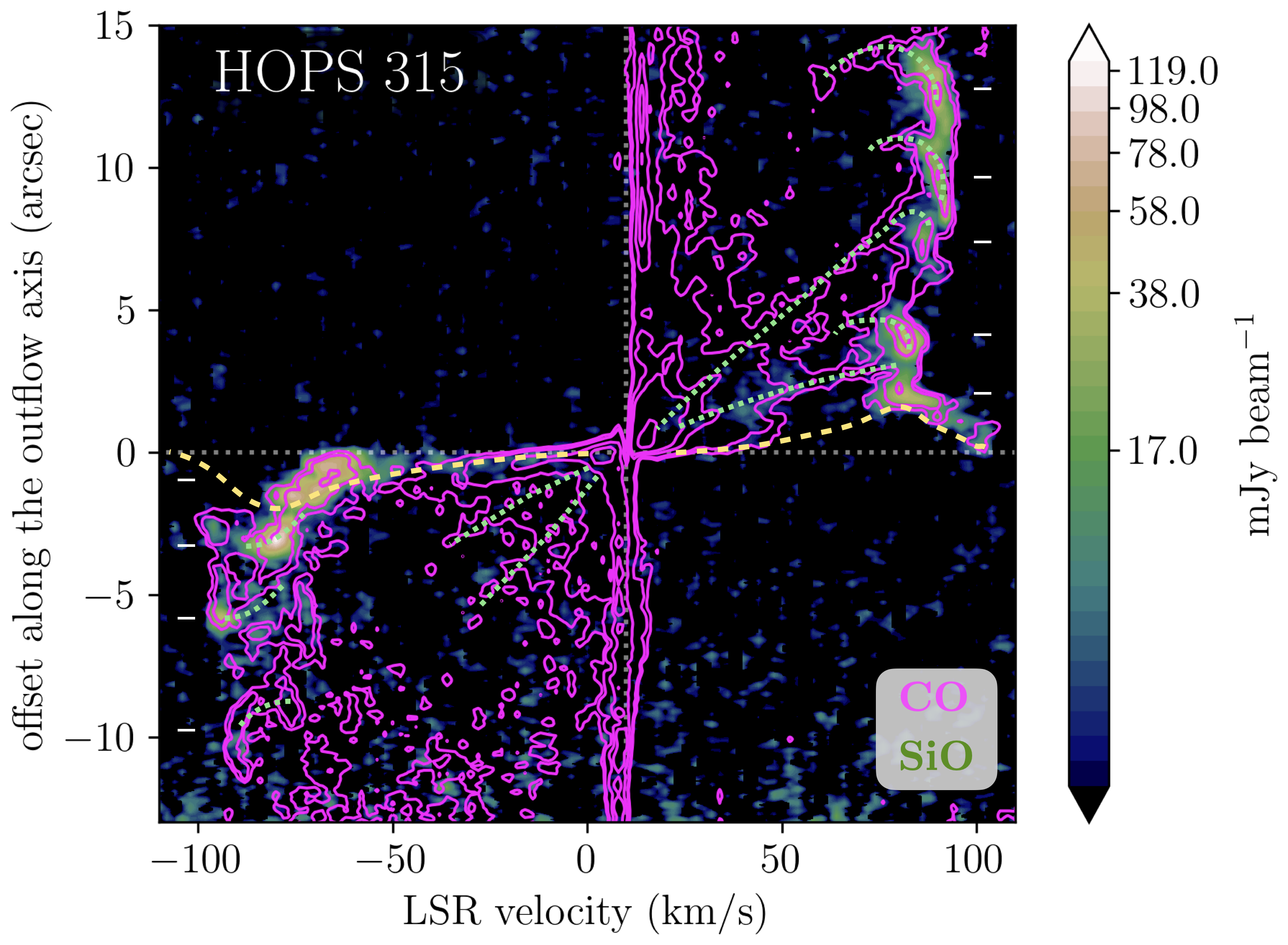}
 \plottwo{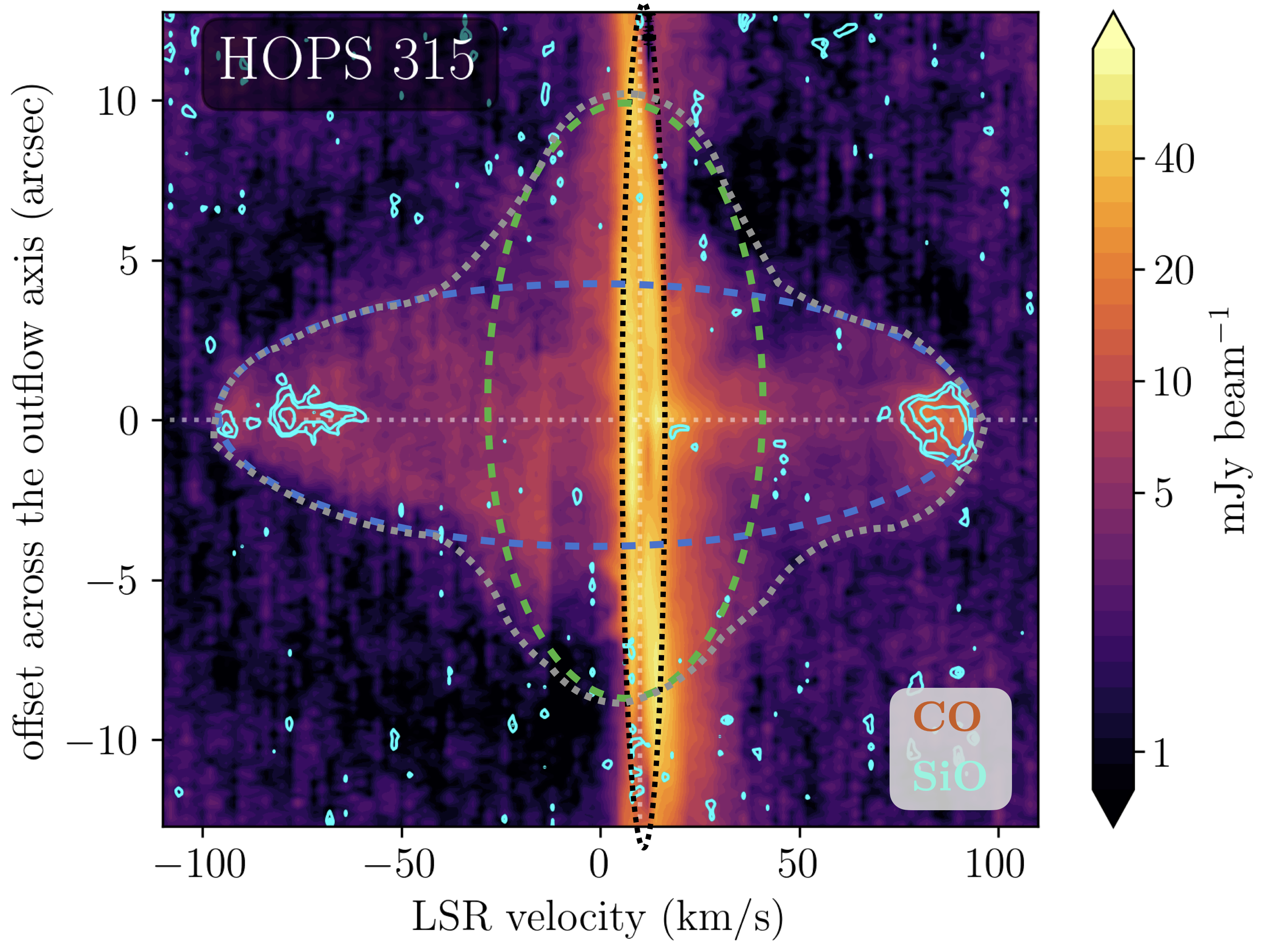}{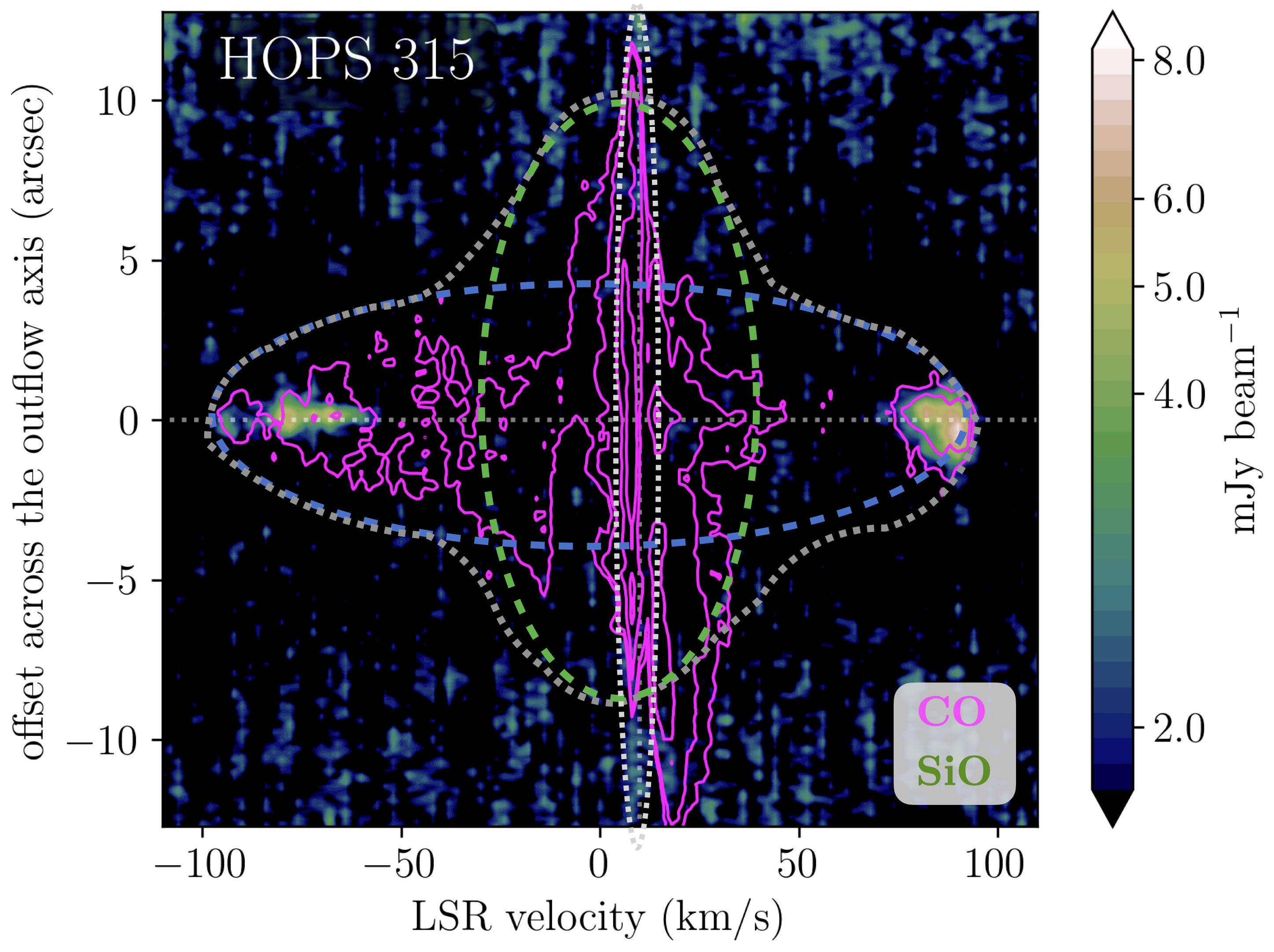}
 \caption{G205.46-14.56S3 (HOPS 315): 
 upper row -- parallel PVDs of CO and SiO along the position angles indicated by thick black lines in Figure \ref{fig:jet-cavity_maps}, with a theoretical slit of $\sim0\farcs6$ width. The position of the source and the value of the systemic velocity in $v_\mathrm{LSR}$ are indicated respectively by the horizontal and vertical dotted lines. The left panel shows CO in the color background and SiO in cyan contours, whereas the right column shows SiO in the color background and CO in magenta contours. On both PVDs, the innermost CO cavity is traced by a yellow dashed line delineating a triangular region at the base of the jet and wide-angle wind. Light-green dotted lines indicate the filamentary structures mainly traced by CO. Positions of knotty SiO structures identified as in, e.g., \citet{jhan2022} are indicated by short horizontal bars.
 lower row -- transverse PVDs perpendicular to the position angles indicated by thick black lines in Figure \ref{fig:jet-cavity_maps}, summing over both the blueshifted and redshifted lobes. The position of the jet axis and the value of the systemic velocity in $v_\mathrm{LSR}$ are indicated respectively by the horizontal and vertical dotted lines. The left column shows CO in background colormap and SiO in cyan contours, and the right column shows SiO in color background and CO in magenta contours. For the CO emission, several ovals are overlaid to show various structures. The overall CO emission can be enclosed within a rounded convex rhombus indicated by the gray dotted curve. Two characteristic ellipses or ovals can be traced within this rhombus (one high-velocity in light blue dashes and one low-velocity in light green dashes). SiO emission can be encompassed by the high-velocity oval close to the two ends of its major axis. A very elongated oval indicates a bright but hollow CO emission around the systemic velocity extended across the outflow lobe.}
 \label{HOPS315_PVDs}
\end{figure*}

\subsubsection{G209.55-19.68S2 (HOPS 10)}\label{subsubsec:HOPS10}
HOPS 10 exhibits asymmetries between the two outflow lobes. 
This is also one of the monopolar SiO jets reported by \citet{jhan2022}, showing predominantly the redshifted SiO jet which extends to more than $15\arcsec$ ($\gtrsim6000\au$) from the source.
The northeastern (redshifted) lobe shows a larger opening angle in CO emission than the southwestern (blueshifted) lobe. Furthermore, the blueshifted CO lobe is brighter than the redshifted lobe. The most prominent features can be seen in the channel maps of CO (Figure \ref{HOPS10_ChMap}). 
For the redshifted side, the lowest velocity channel adjacent to the systemic velocity shows a limb-brightened shell filled with diffuse CO emission. With increasing velocity, the shape becomes rounded with a bubble-like structure, and then decreases in spatial width until the highest velocities (associated with the HVC) in which SiO emission also appears.
At various velocity channels, nested shell-like structures with different length-to-width ratios can be seen. These structures also manifest in the PVDs, corresponding to the CO emission existing within the region between the LV and EHV ranges (Figure \ref{HOPS10_PVDs}). The properties of the HOPS 10 CO outflows were previously studied by \citet{hsieh2023}. 

From the PVDs parallel to the jet axis (upper row of Figure \ref{HOPS10_PVDs}), CO emission can be seen filling the region between the LV and EHV ranges. For the redshifted lobe, at close to the source, a triangular-shaped CO void can be seen. The SiO emission starts to emerge only after the tip of this triangular cavity. The SiO knotty structures are delineated as white dotted curves in the parallel PVDs. For the blue-shifted lobe, the velocity extension of the CO emission is much less, leaving only a faint trace of the spindle emission extending toward the opposite direction to that of the redshifted part.

For the transverse PVDs (lower row of Figure \ref{HOPS10_PVDs}), the CO emission approximates a half oval, essentially limited to the redshifted side. As shown in Figure \ref{HOPS10_ChMap}, the blueshifted CO emission extends only to $v_\mathrm{LSR}\approx-15\kms$ and no faster CO nor SiO emission can be found as opposed to the redshifted lobe, which is consistent with this appearance. At the redshifted high-velocity end of the oval, a bright knot can be seen in both CO and SiO emission, tracing the EHV emission. A very elongated oval is drawn in black dots that extends further across the lobe and has a very low velocity close to the systemic velocity.

\begin{figure*}
 \begin{center}
 \includegraphics[angle=-90,scale=0.7]{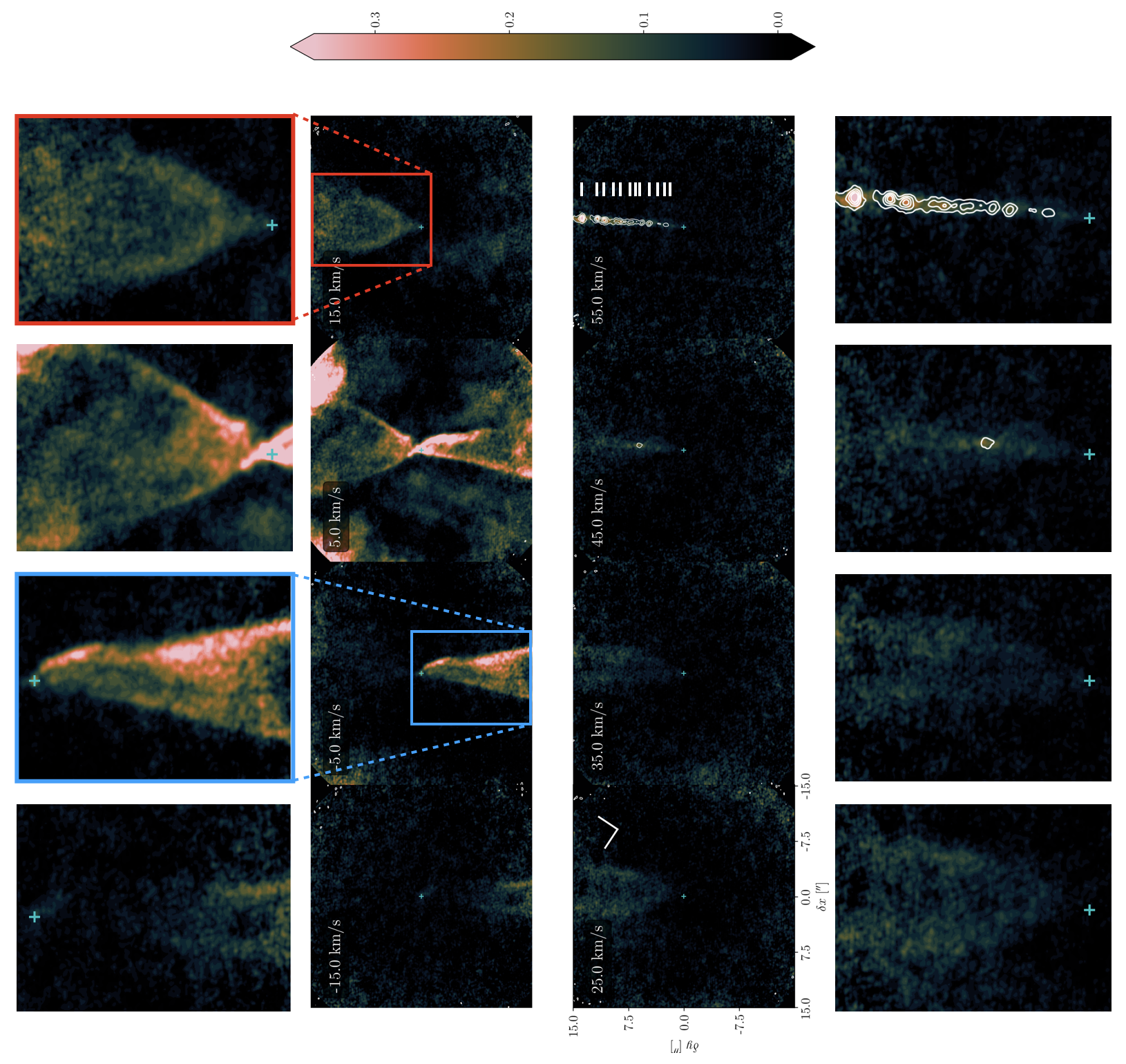}
 \caption{G209.55-19.68S2 (HOPS 10): integrated channel maps with $\sim10\kms$ width per channel shown in the central two rows. The central velocity of each channel is shown in $v_\mathrm{LSR}$. The maps have been rotated to vertically align the redshifted outflow axis upwards. North--east compass bars are given in the lower-left panel. The maps show $^{12}$CO in color background and SiO in white contours in units of Jy\,beam$^{-1}$$\kms$. Positions of high-velocity SiO blobs identified in \citet{jhan2022} are indicated by short horizontal bars. The uppermost and lowermost rows show zoomed-in views of the blueshifted and redshifted lobes with field of views indicated by the blue and red rectangles, respectively. In each of the panels, the nominal position of the star is drawn as a green cross.}\label{HOPS10_ChMap}
 \end{center}
\end{figure*}
\begin{figure*}
 \plottwo{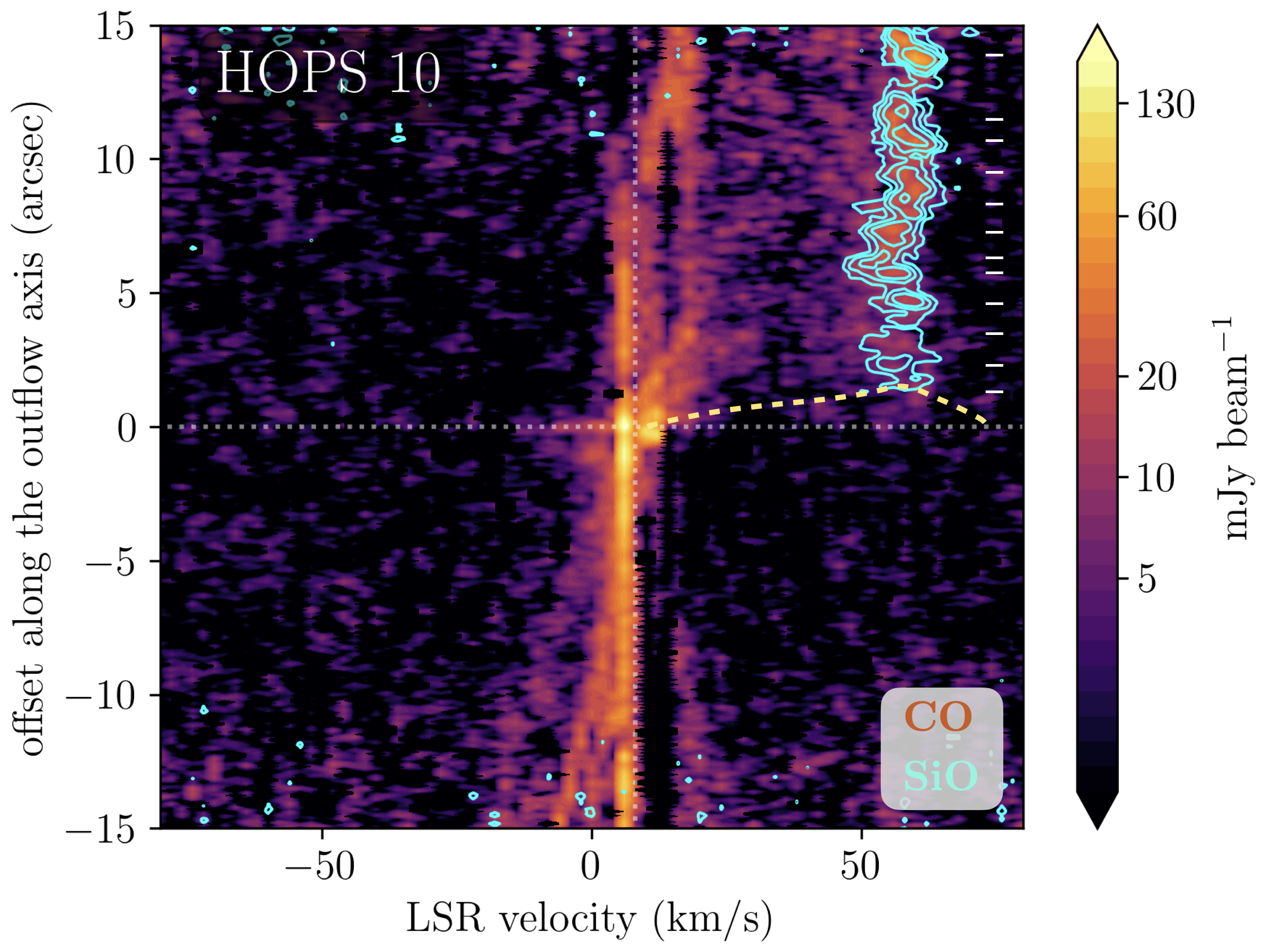}{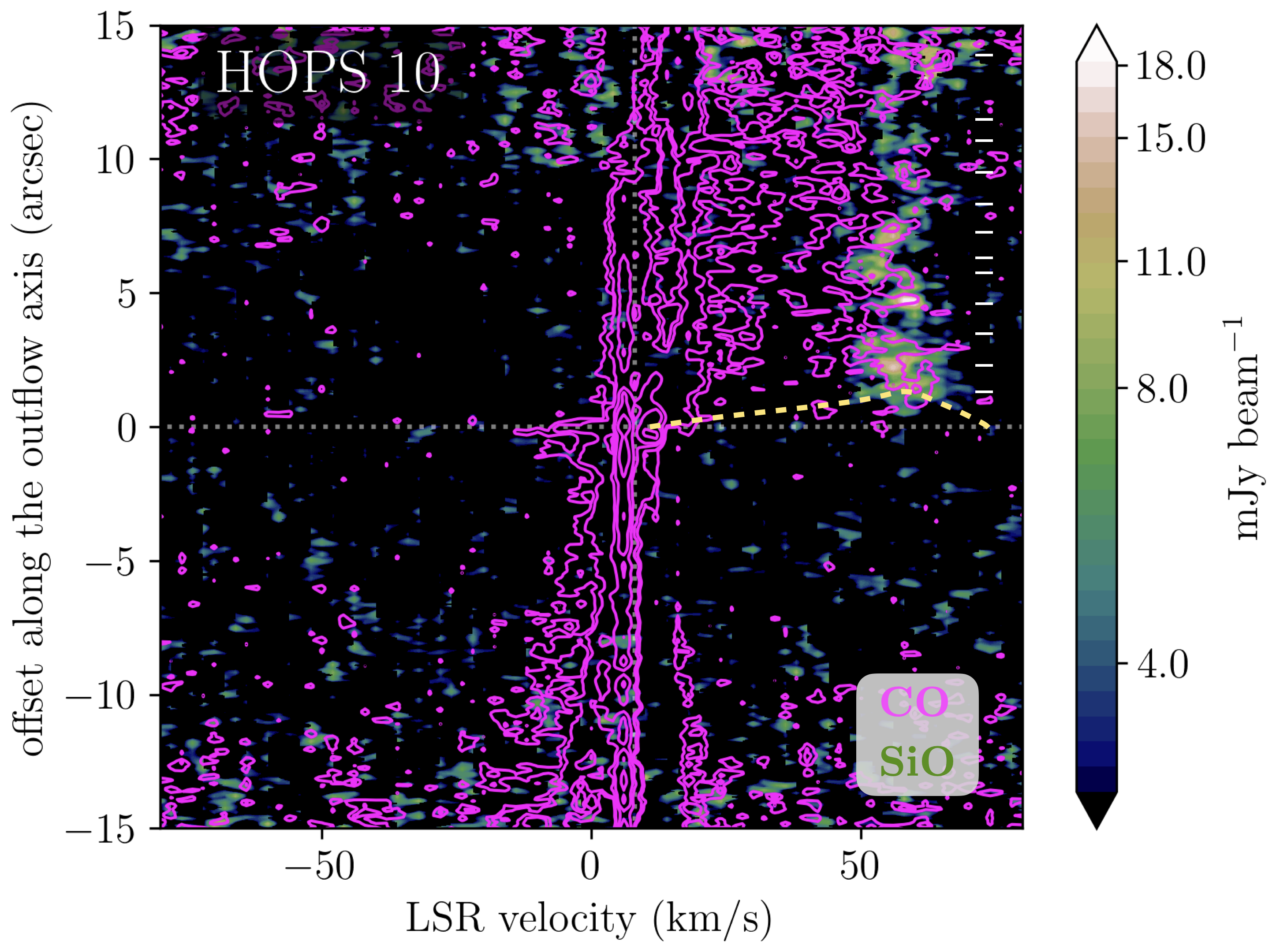}
 \plottwo{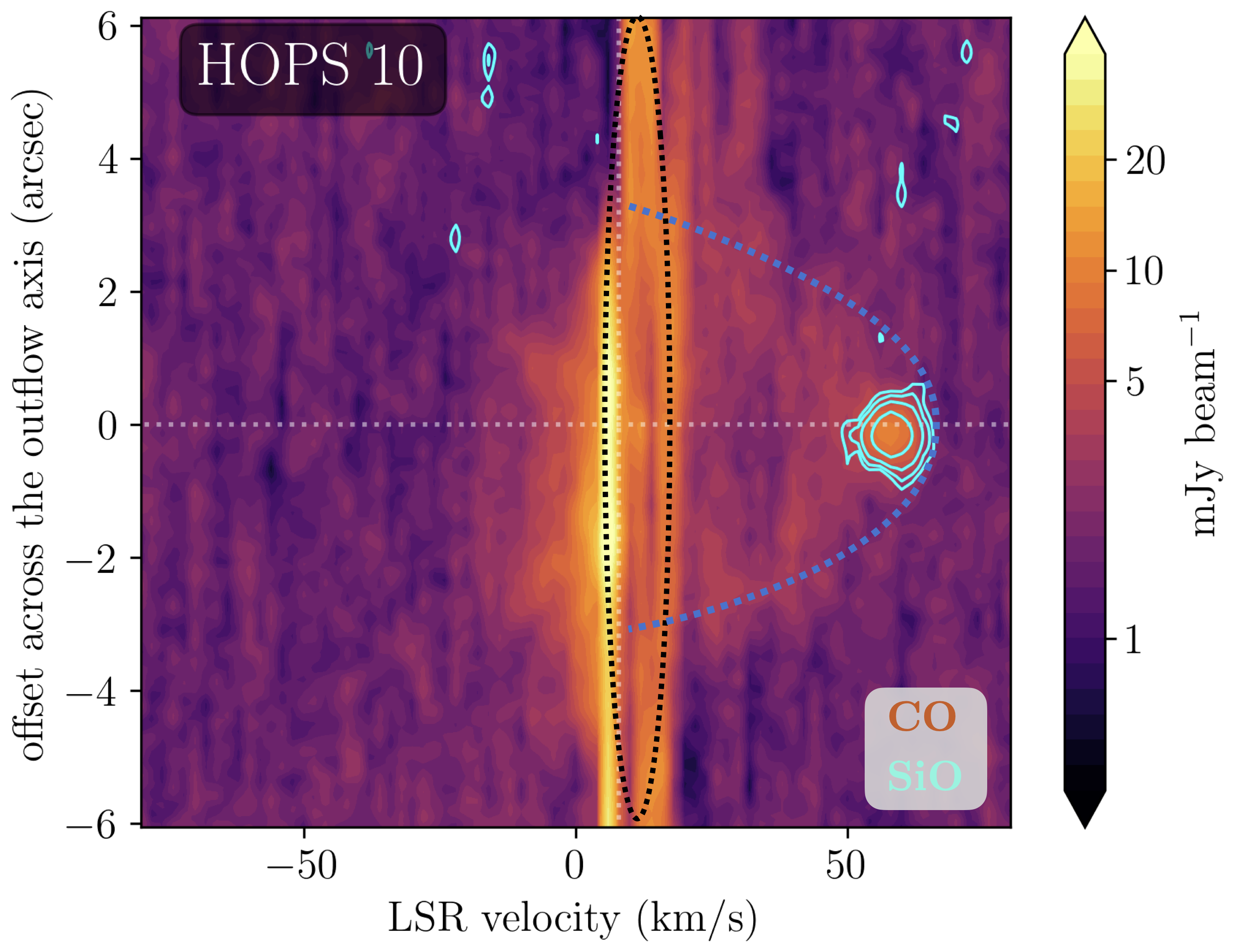}{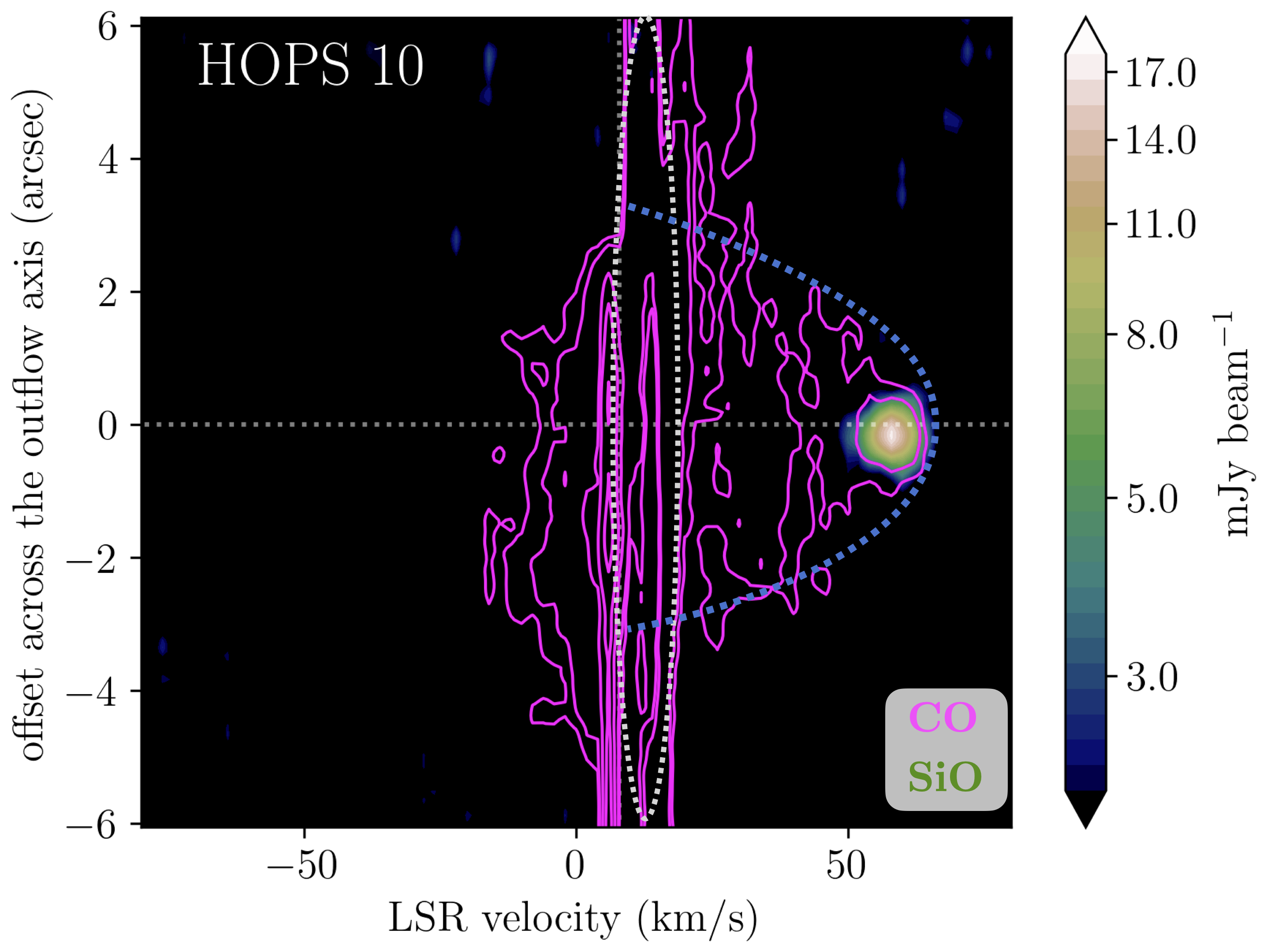}
 \caption{G209.55-19.68S2 (HOPS 10): 
 upper row -- parallel PVDs of CO and SiO along the position angles indicated by thick black lines in Figure \ref{fig:jet-cavity_maps}, with a theoretical slit of $\sim0\farcs6$ width. The position of the source and the value of the systemic velocity in $v_\mathrm{LSR}$ are indicated respectively by the horizontal and vertical dotted lines. The left panel shows CO in the color background and SiO in cyan contours, whereas the right column shows SiO in the color background and CO in magenta contours. On both PVDs, the innermost CO cavity is traced by a yellow dashed line delineating a triangular region at the base of the jet and wide-angle wind. Positions of knotty SiO structures identified as in, e.g., \citet{jhan2022} are indicated by short horizontal bars.
 lower row -- transverse PVDs perpendicular to the position angles indicated by thick black lines in Figure \ref{fig:jet-cavity_maps}, summing over both the blueshifted and redshifted lobes. The position of the jet axis and the value of the systemic velocity in $v_\mathrm{LSR}$ are indicated respectively by the horizontal and vertical dotted lines. The left column shows CO in background colormap and SiO in cyan contours, and the right column shows SiO in color background and CO in magenta contours. For CO, the redshifted lobe dominates the emission and thus only a half oval shape can be traced (by the light-blue dashed curve). SiO emission locates at the jet velocity as shown in the parallel PVDs and is encompassed by the half oval close to its maximum velocity. A bright but hollow CO emission around the systemic velocity extended across the outflow lobe is indicated by a very vertically elongated oval.}
\label{HOPS10_PVDs}
\end{figure*}

\subsubsection{G203.21-11.20W2 (G203W2)}\label{subsubsec:G203W2}
G203W2 has been identified as a hot corino source \citep{hsu2022}. It shows a bipolar outflow in CO (Figure \ref{G203W2_ChMap}) that appears symmetric between the two lobes in terms of morphology. For both lobes at close to the systemic velocity of $v_{\rm LSR} = 10\kms$, the CO emission shows the appearance of limb-brightened shells with an opening angle of $\sim35^\circ$. Reaching $\pm15$ to $\pm25\kms$ from the systemic velocity, CO emission narrows down to $\lesssim10^\circ$ and appears knotty close to the base and with a bubble-like structure downstream, traced by the cyan dotted lines. SiO, on the other hand, shows an asymmetric appearance with stronger intensities in the redshifted (southern) lobe. The redshifted SiO outflow appears jet-like and knotty. The blueshifted SiO outflow can only be seen as a compact blob close to the source.

The asymmetric kinematic structures of SiO can be observed in both the parallel and transverse PVDs (Figure \ref{G203W2_PVDs}). The redshifted SiO has a centroid velocity of $\sim35\kms$ with respect to $v_\mathrm{sys}$ and extends to $\sim10\arcsec$ from the source. The parallel PVDs for the redshifted SiO and CO outflow were also reported by \citet{jhan2022}. 
The compact blueshifted SiO knot shows a large line width of $>30\kms$ at the jet base with a centroid velocity of $\sim-25\kms$ with respect to $v_\mathrm{sys}$. The large line width at the base suggests a wide-angle wind. The density and excitation asymmetry between the two outflow lobes may be the main reason for the SiO intensity asymmetry. 

For the brighter redshifted lobe, the CO parallel PVD (Figure \ref{G203W2_PVDs}, upper left) shows filamentary threads connecting the LV to EHV regions as traced by light-green dotted lines, filling in a roughly rectangular region in the PV space. A triangular-shaped CO void can be seen close to the source position, delineated by yellow dashed lines. SiO emission emerges only beyond the apex of the triangular void (Figure \ref{G203W2_PVDs}, upper right). The filamentary thread} is seen to connect to the knotty EHV structure in CO and SiO emission. 
Beyond the triangular void, the SiO emission shows an overall head--tail velocity slope structure, resembling the CO and SiO patterns 
found in IRAS 04166+2706 \citep{santiago-garcia2009,wang2015}.  A discontinuous change of velocity and intensity can be seen in the before the SiO emission at $\sim8\arcsec$ and $\sim12\arcsec$ where the bubble/shell structures appear. 

The CO transverse PVD (Figure \ref{G203W2_PVDs}, lower left) shows emission with the shape of a concave rhombus (gray dotted line), roughly centered on the systemic velocity and source position. Two characteristic ellipses or ovals can be traced within this rhombus (one high-velocity in light blue dashes, and one low-velocity in light green dashes). An additional bright oval (black dots) corresponds to even smaller line-of-sight velocities close to the systemic velocity, and extends to larger positions.
The SiO emission appears as a bright knot near the $+v$ end of the high-velocity oval major axis. A dimmer knot exists at the $-v$ end.
The redshifted side has a brighter SiO knot at $\sim35\kms$ with respect to $v_\mathrm{sys}$, and the blueshifted side has a dimmer knot at $\sim-35\kms$ with respect to $v_\mathrm{sys}$, tracing the inner axial portion of G203W2.

\begin{figure*}
 \begin{center}
 \includegraphics[angle=-90,scale=0.7]{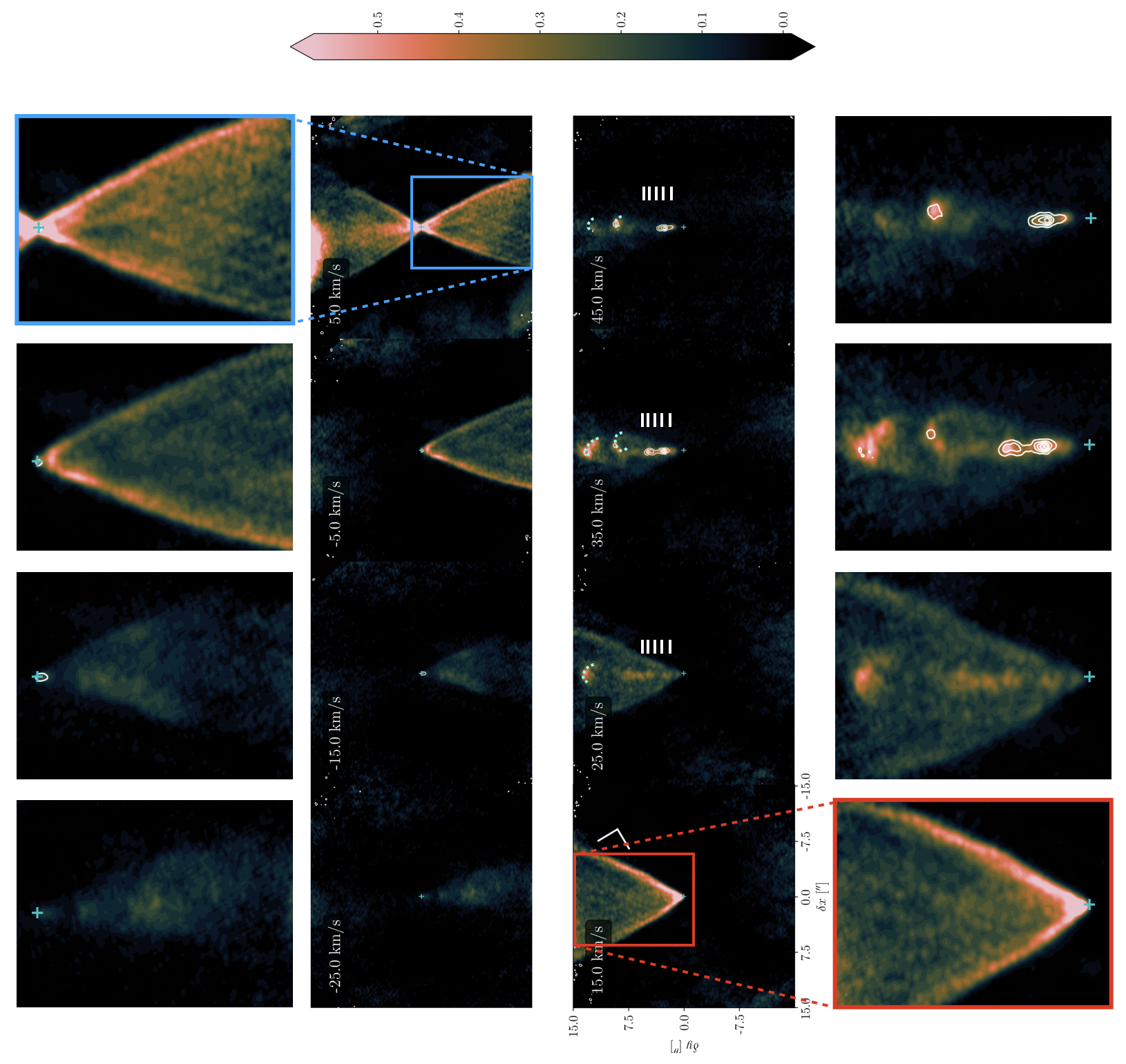}
 \caption{G203.21-11.20W2 (G203W2): 
 integrated channel maps with $\sim10\kms$ width per channel shown in the central two rows. The central velocity of each channel is shown in $v_\mathrm{LSR}$. The maps have been rotated so that the redshifted outflow axis is upwards with north--east compass bars in the lower-left panel. The maps show $^{12}$CO in color maps and SiO in white contours in units of Jy\,beam$^{-1}\kms$. 
 Positions of high-velocity SiO blobs identified in \citet{jhan2022} are indicated by short horizontal bars, and arc-like and bubble-like CO structures are delineated by cyan dotted lines. The uppermost and lowermost rows show zoomed-in views of the blueshifted and redshifted lobes with field of views indicated by the blue and red rectangles, respectively. In each of the panels, the nominal position of the star is drawn as a green cross.}  \label{G203W2_ChMap}
 \end{center}
\end{figure*}
\begin{figure*}
 \plottwo{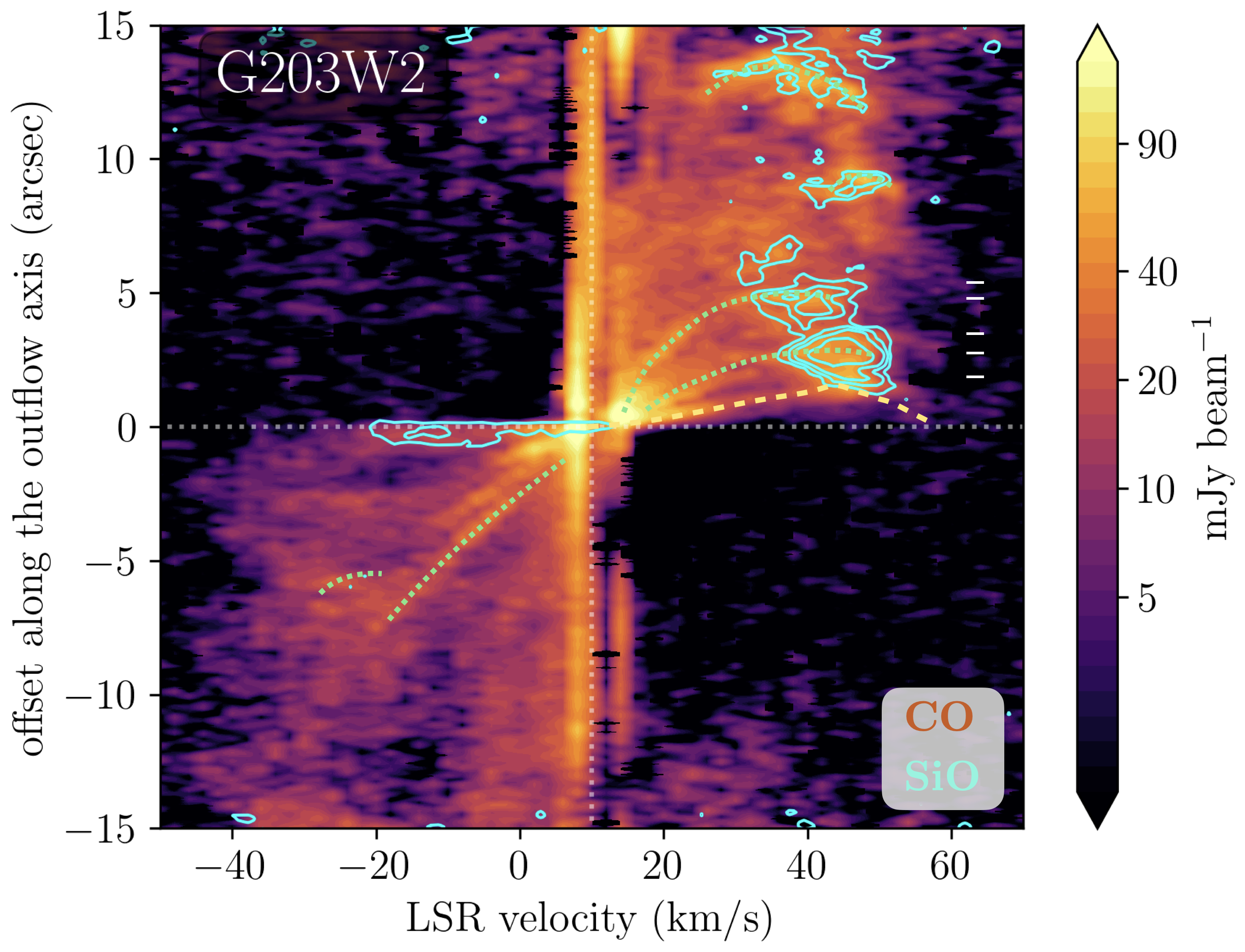}{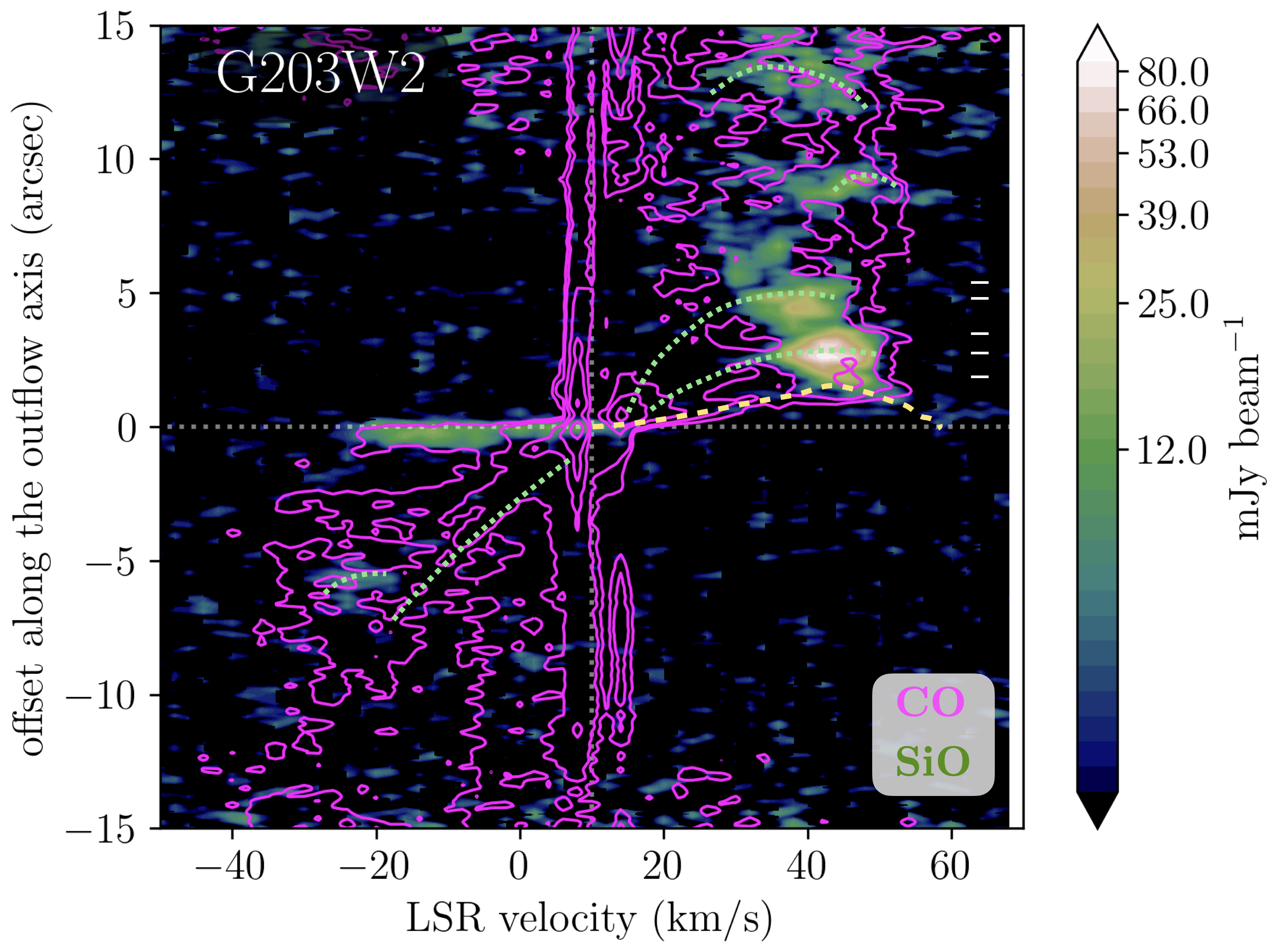}
 \plottwo{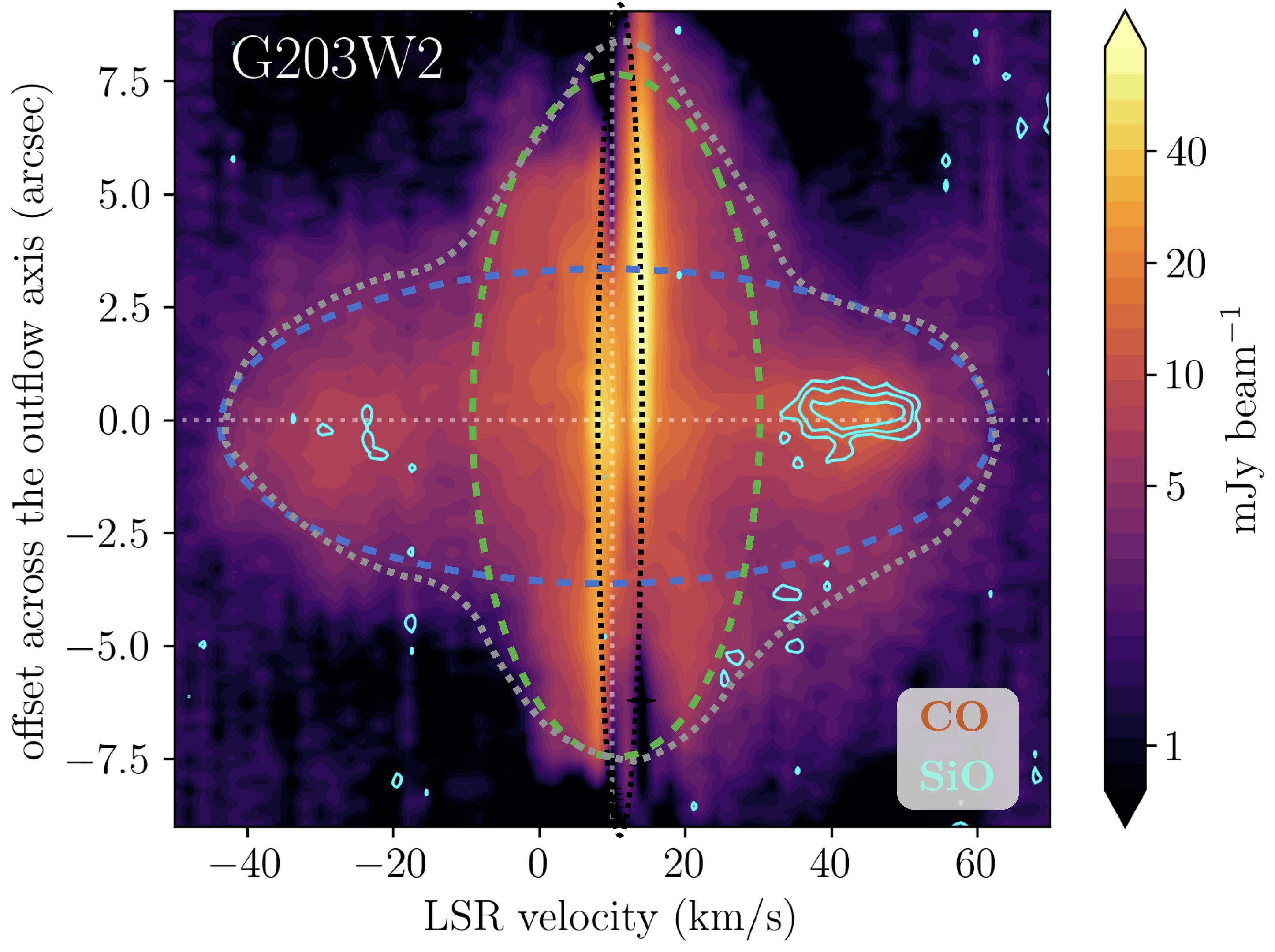}{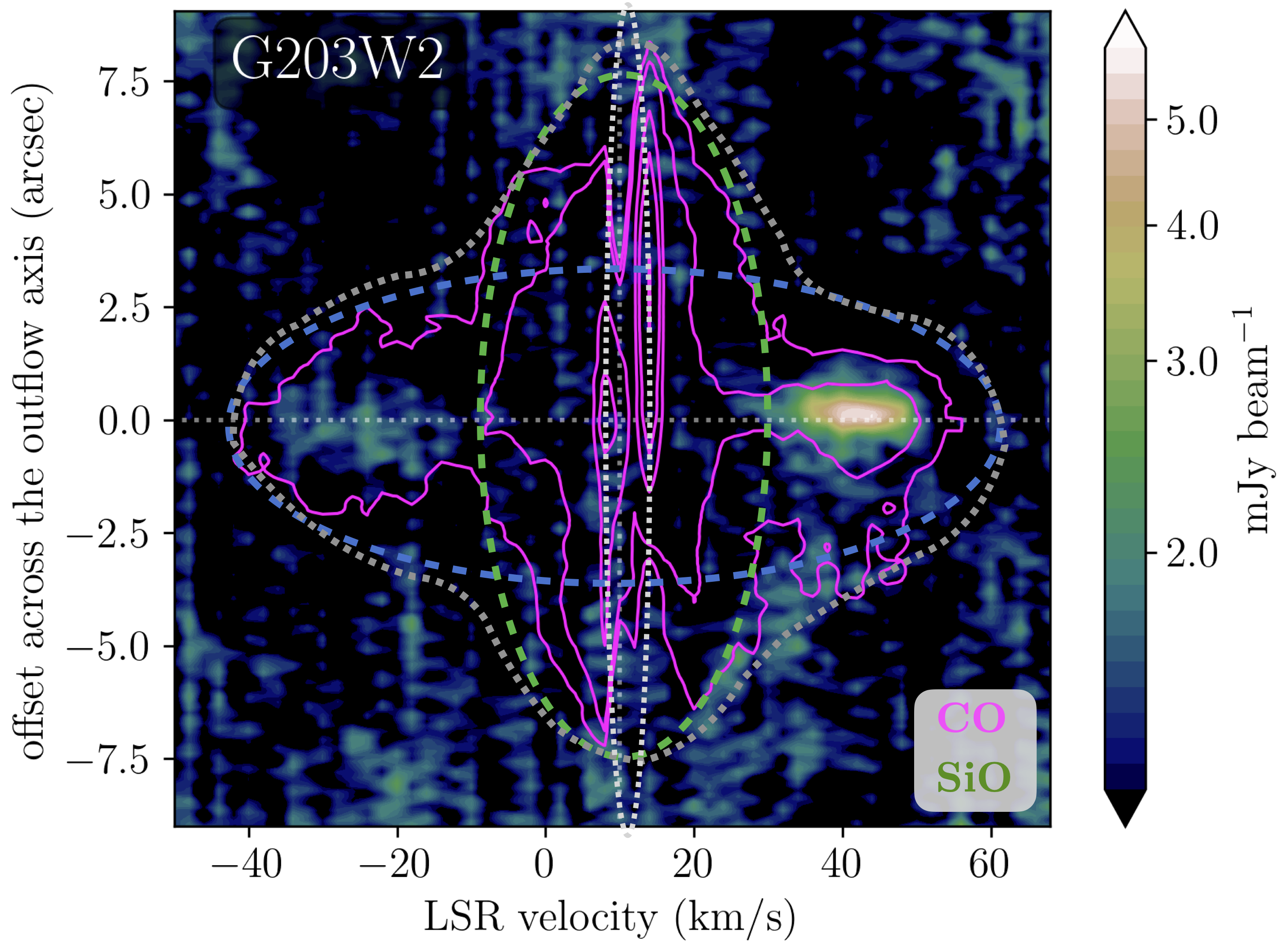}
 \caption{G203.21-11.20W2 (G203W2):
 upper row -- parallel PVDs of CO and SiO along the position angles indicated by thick black lines in Figure \ref{fig:jet-cavity_maps}, with a theoretical slit of $\sim0\farcs6$ width. The position of the source and the value of the systemic velocity in $v_\mathrm{LSR}$ are indicated respectively by the horizontal and vertical dotted lines. The left panel shows CO in the color background and SiO in cyan contours, whereas the right column shows SiO in the color background and CO in magenta contours. On both PVDs, the innermost CO cavity is traced by a yellow dashed line delineating a triangular region at the base of the jet and wide-angle wind. The filamentary structures mainly traced by CO are indicated by light-green dotted lines. Positions of knotty SiO structures identified as in, e.g., \citet{jhan2022} are indicated by short horizontal bars.
 lower row -- transverse PVDs perpendicular to the position angles indicated by thick black lines in Figure \ref{fig:jet-cavity_maps}, summing over both the blueshifted and redshifted lobes. The position of the jet axis and the value of the systemic velocity in $v_\mathrm{LSR}$ are indicated respectively by the horizontal and vertical dotted lines. The left column shows CO in background colormap and SiO in cyan contours, and the right column shows SiO in color background and CO in magenta contours. For the CO emission, several ovals are overlaid to show various structures. The overall CO emission can be enclosed within a rounded convex rhombus indicated by the gray dotted curve. Two characteristic ellipses or ovals can be traced within this rhombus (one high-velocity in light-blue dashes and one low-velocity in light-green dashes). SiO emission can be encompassed by the high-velocity oval close to the two ends of its major axis. A very elongated oval indicates a bright but hollow CO emission around the systemic velocity extended across the outflow lobe.}
 \label{G203W2_PVDs}
\end{figure*}

\subsubsection{G205.46-14.56S1 (HOPS 358)}\label{subsubsec:HOPS358}
\begin{figure*}
 \begin{center}
 \includegraphics[angle=-90,scale=0.7]{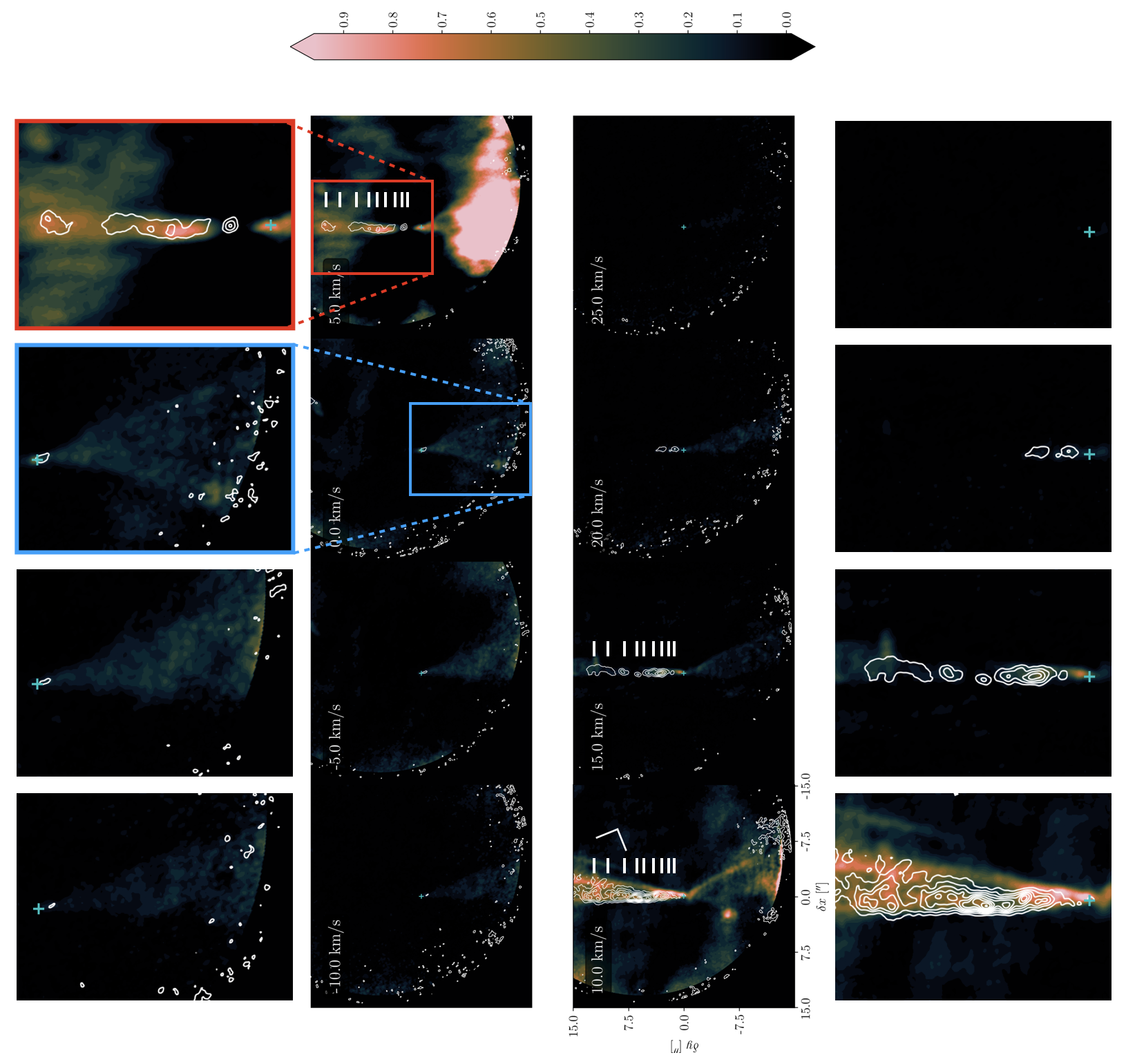}
 \caption{G205.46-14.56S1 (HOPS 358): integrated channel maps with $\sim10\kms$ width per channel shown in the central two rows. The central velocity of each channel is shown in $v_\mathrm{LSR}$. The maps have been rotated so that the redshifted outflow axis is upwards with north--east compass bars in the lower-left panel. The maps show $^{12}$CO in color maps and SiO in white contours in units of Jy\,beam$^{-1}\kms$. 
 Positions of high-velocity SiO blobs identified in \citet{jhan2022} are indicated by short horizontal bars, and arc-like and bubble-like CO structures are delineated by cyan dotted lines. The uppermost and lowermost rows show zoomed-in views of the blueshifted and redshifted lobes with field of views indicated by the blue and red rectangles, respectively. In each of the panels, the nominal position of the star is drawn as a green cross.} 
 \label{HOPS358_ChMap}
 \end{center}
\end{figure*}

This source has been reported by \citet{jhan2022} as G205.46-14.56S1\_A. It corresponds to HOPS 358 \citep{furlan2016} and shows a temporally-changing sub-mm continuum peak flux \citep{lee2021_Transient,mairs2024}. It is also known as a hot corino source \citep{hsu2022}. The outflow shows an obvious asymmetry between the blueshifted and redshifted lobes. The SiO emission is predominantly strong in the redshifted lobe, mainly tracing the molecular jet. Both CO and SiO show narrow cone-like structures (Figure \ref{HOPS358_ChMap}). The SiO jet is characterized as a series of knots. 

\begin{figure*}
 \plottwo{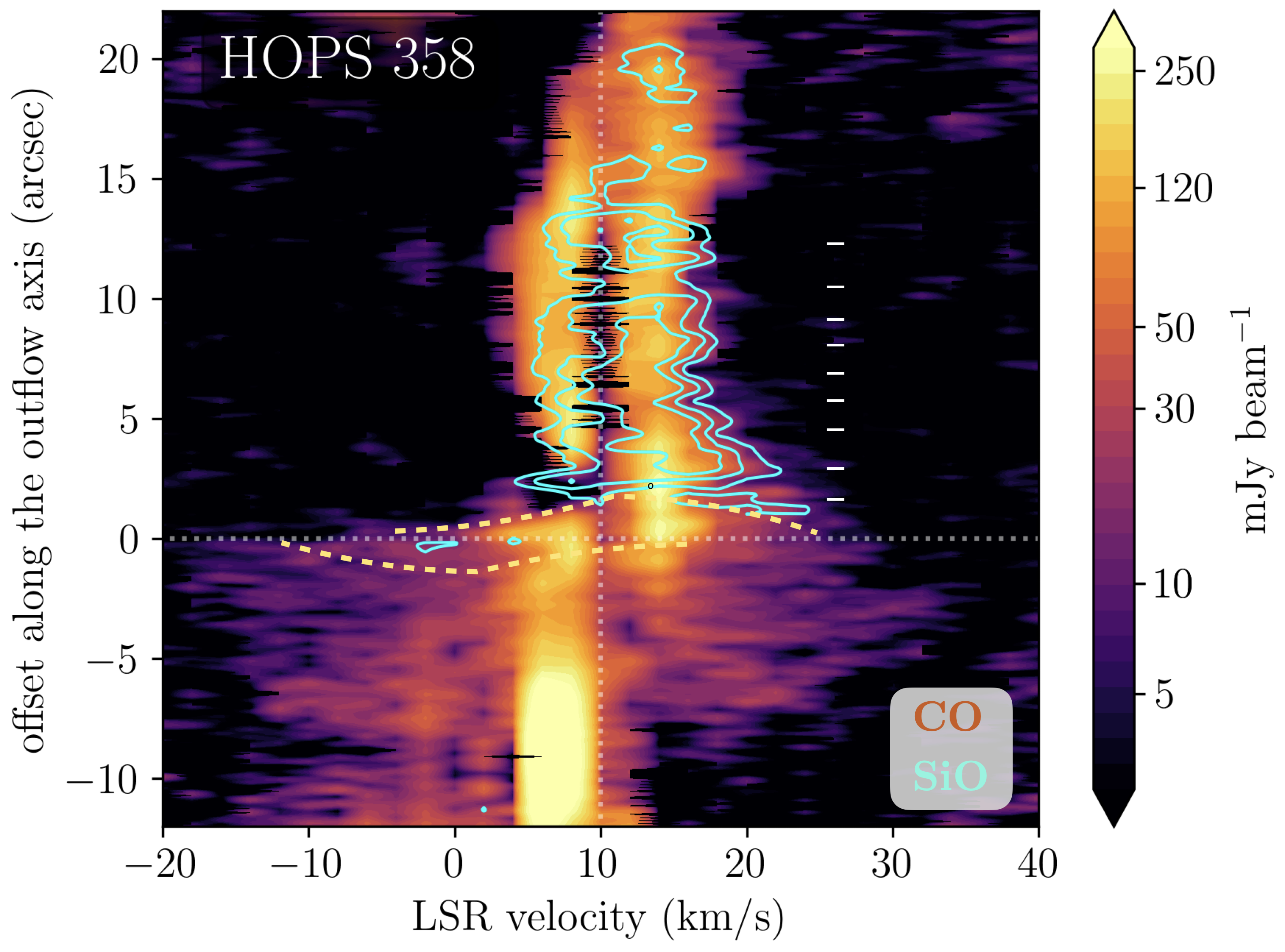}{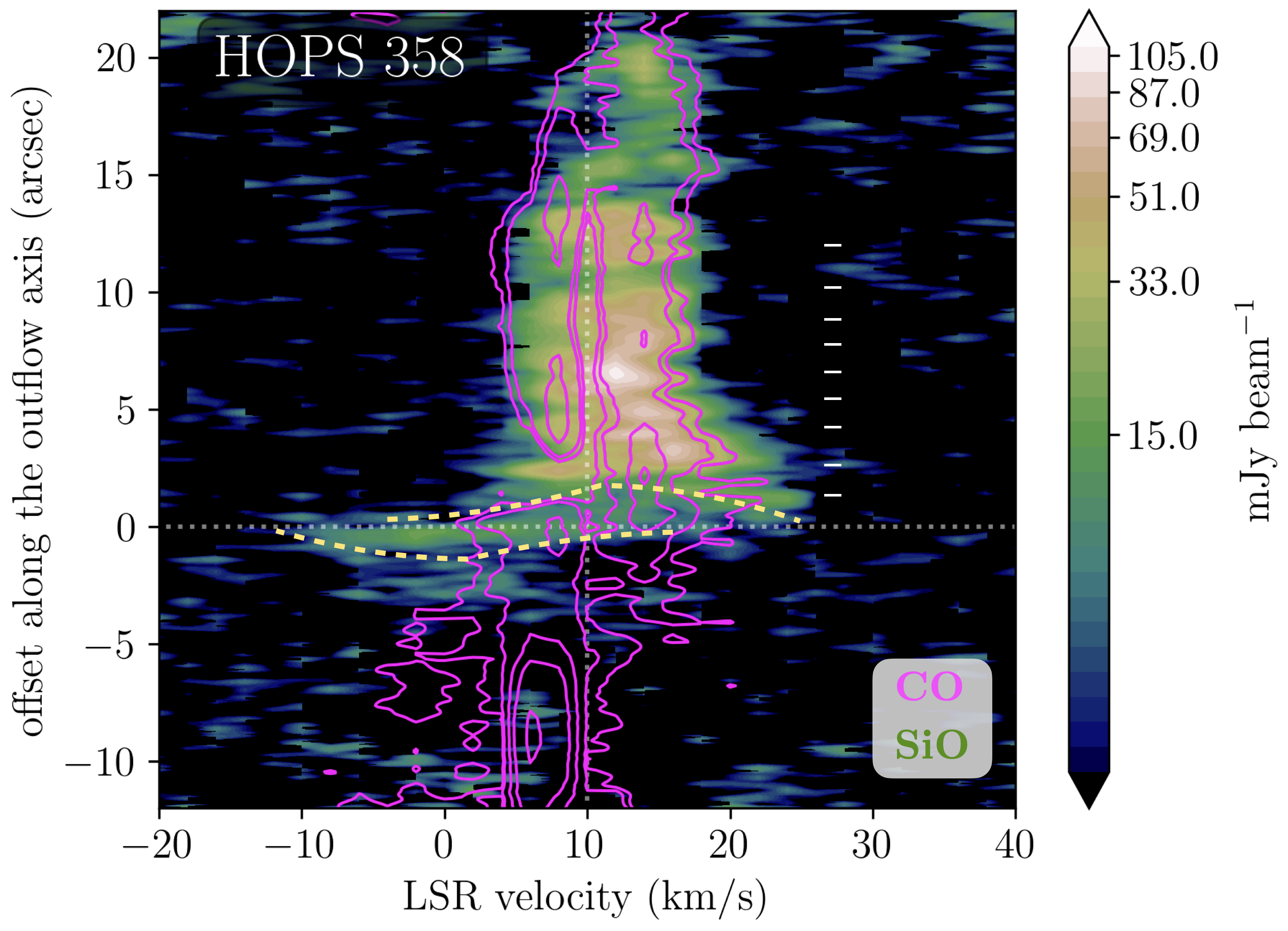}
 \plottwo{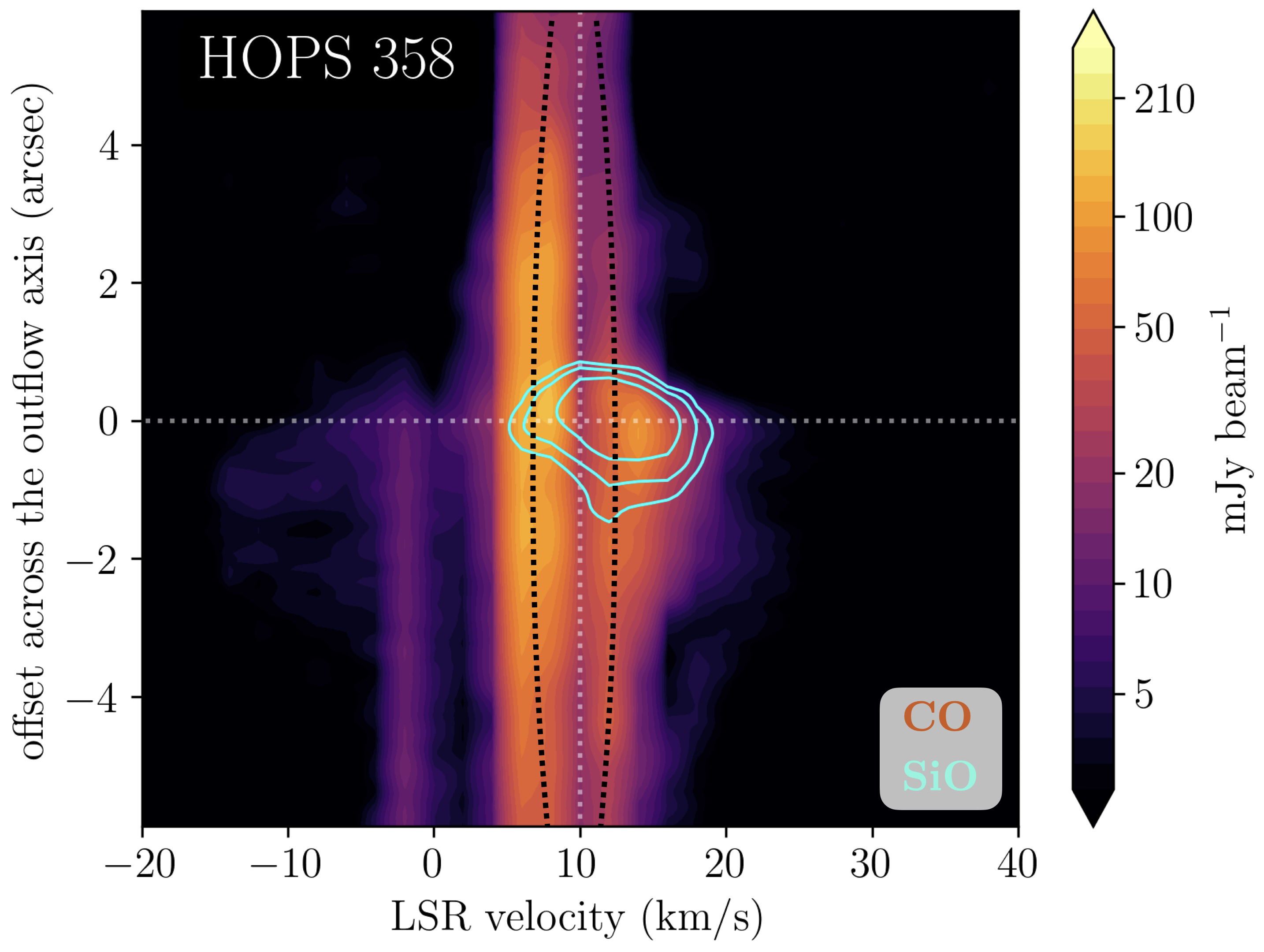}{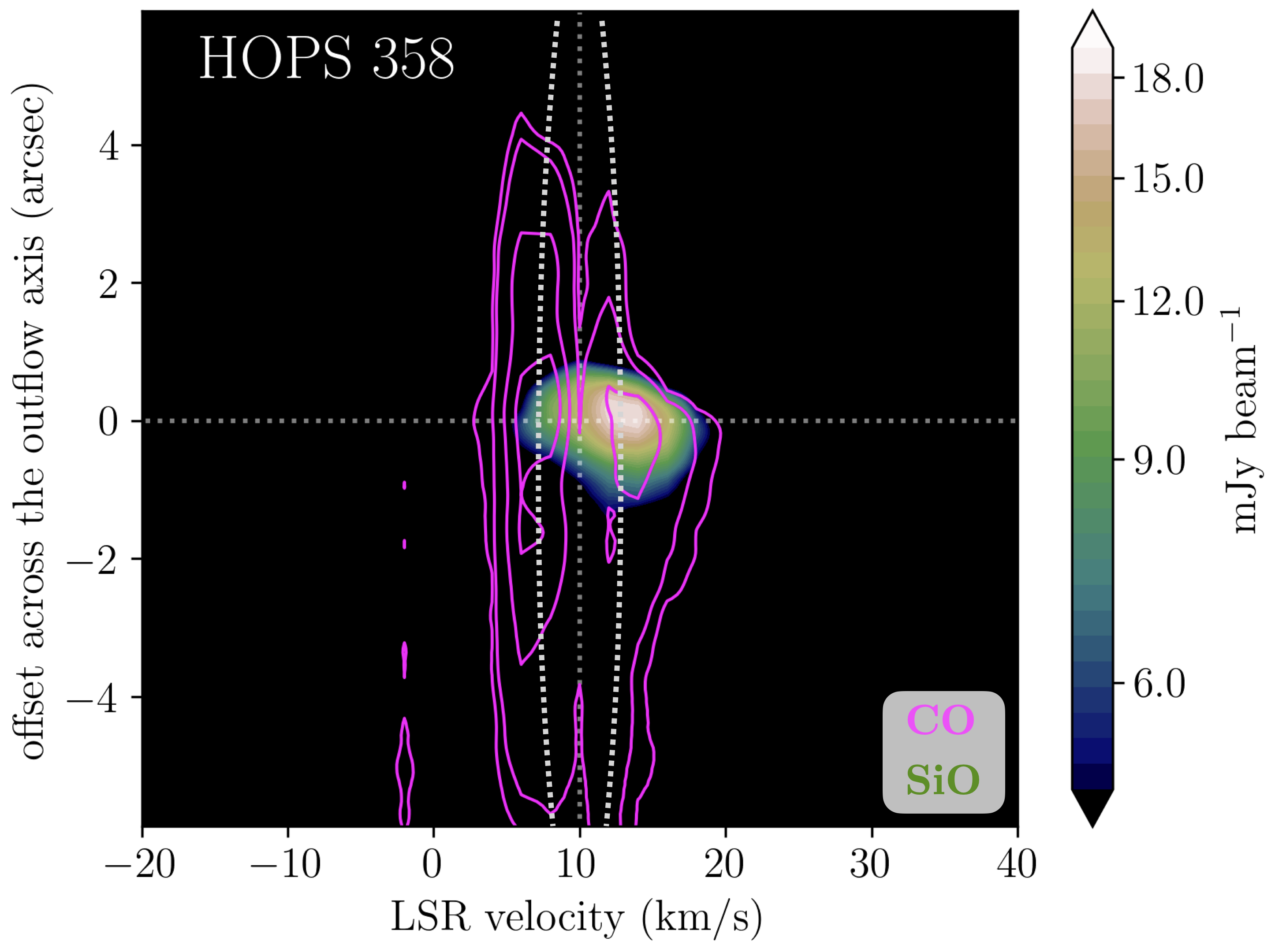}
 \caption{G205.46-14.56S1 (HOPS 358):
 upper row -- parallel PVDs of CO and SiO along the position angles indicated by thick black lines in Figure \ref{fig:jet-cavity_maps}, with a theoretical slit of $\sim0\farcs6$ width. The position of the source and the value of the systemic velocity in $v_\mathrm{LSR}$ are indicated respectively by the horizontal and vertical dotted lines. The left panel shows CO in the color background and SiO in cyan contours, whereas the right column shows SiO in the color background and CO in magenta contours. On both PVDs, the innermost CO cavity is traced by a yellow dashed line delineating a triangular region at the base of the jet and wide-angle wind. Positions of knotty SiO structures identified as in, e.g., \citet{jhan2022} are indicated by short horizontal bars.
 lower row -- transverse PVDs perpendicular to the position angles indicated by thick black lines in Figure \ref{fig:jet-cavity_maps}, summing over both the blueshifted and redshifted lobes. The position of the jet axis and the value of the systemic velocity in $v_\mathrm{LSR}$ are indicated respectively by the horizontal and vertical dotted lines. The left column shows CO in background colormap and SiO in cyan contours, and the right column shows SiO in color background and CO in magenta contours. For CO, the emission is dominated by a bright but hollow CO emission around the systemic velocity extended across the outflow, indicating a bright surrounding environment. It is indicated by a very vertically elongated oval. SiO emission locates at the jet velocity as shown in the parallel PVDs, which is close to the systemic velocity due to its inclination close to edge-on. The rest of CO emission is overshadowed by the ambient material contribution and thus no proper estimate of the coverage can be made.}
\label{HOPS358_PVDs}
\end{figure*}

In the parallel PVDs (upper right of Figure \ref{HOPS358_PVDs}), the SiO jet is shown close to the systemic velocity due to its close to edge-on inclination. The base of the SiO jet has a large line width and the consecutive knots reveal a saw-tooth pattern. The saw-tooth pattern spans from $v_{\rm LSR} \approx 7\kms$ to $\approx 20\kms$ ($v_{\rm sys}\approx10\kms$). On the other hand, the CO emission is affected by the bright emission most likely from the ambient material which is also centered at the systemic velocity.
In the transverse PVDs, the CO emission can be seen to be dominated by the bright emission with low velocities close to the systemic velocity, which is traced by an elongated oval with black dots. A trace of another oval structure extending toward the higher velocities may be seen. The SiO emission appears as a featureless blob, centered at the projected jet velocity, close to the systemic velocity, on the jet axis. This lack of any spatial distribution indicates that the bright SiO material is associated with the inner axial jet.

\section{Observed Outflow Features in the Context of Theoretical Models} \label{sec:discussion}

We discuss the current observational findings and, by analyzing against a range of theoretical scenarios, note the need for sophisticated models such as the magnetized bubble structures advanced by \citet{shang2020,shang_PII}.
Next, we separately investigate the significance of the reverse shock cavities, the nested kinematic features, the origins of SiO emission, and apparent episodicity. This is followed by consideration of the conundrums posed by alternate conventional scenarios, including difficulties invoking slow molecular winds.

\subsection{Historically Recognized Features} \label{subsec:history}

We start with the features that constitute the historical understanding of jets and outflows. 
These prominent features can be identified and compared in the current sources and the respective theoretical interpretations. 

\citet{shu1991} advanced the first framework of a momentum-conserving hydrodynamic wind-driven thin shell. The model relies on two key parameters: the angular distribution of both the wind momentum and the ambient density, labeled as $P$ and $Q$. Appropriate specification of $P$ and $Q$ required detailed wind and ambient medium theories. The X-wind model asymptotics \citep{shu1995} provided a wind factor $P\propto \sin^{-2}\theta$, and the singular toroids by \citet{li1996b} provided an ambient ratio factor $Q(\theta)\equiv R_n(\theta)$ depending on a toroid opening parameter $0<n<\infty$.
Both $P$ and $Q$ are suitably anisotropic and thus able to produce a thin shell with a bipolar outflow appearance. For the usual range of $n$ values adopted by \citet{shang2006,shang2020}, the resulting spindle-shaped lobes describe the hydrodynamical momentum-conserving thin-shell outflows, e.g.,\ Figure 9 of \citet{shang_PII}.

Similarly, \citet{lee2000,lee2001} examined analytical models, carried out pure hydrodynamical simulations of wind-driven shells and jet-driven bow shocks, and found valuable empirical fits which were useful for comparison against the features being observed at the time.
\citet{lee2000} Equation (1) and schematic Figure 21 fit their wind-driven outflow shapes at the base to a parabolic function.
Explicitly, \citet{lee2000,lee2001} adopted the specific $\sin^2\theta/r^2$ ambient density profile, which is similar to $n = 2$ in Figure 5a of \cite{li1996b}. 

These valuable historical models shared the common limitation of considering only thin shells and not directly accounting for wind and ambient medium magnetization. As the observational evidence accumulates, both jet- and wind-driven features are required to understand the phenomena. Advancement to a unified model containing jet-driven and wind-driven features \citep{shang2006} is a theoretical milestone of the interpretation. 

\citet{shang2006} numerically extended the \citet{shu1991} model, applying the \citet{li1996b} singular isothermal toroids as the surrounding environment. The interaction between the magnetized wind containing an asymptotically cylindrically stratified density and ambient toroids with various flattening due to differences in the extent of poloidal magnetic support leads to ``spindle-like'' outflows shaped with a variety of length-to-width ratios. The un-impeded faster shell material travels outward along the least dense regions of the prenatal cloud core, while the stagnant base anchors the much denser accreting and infalling pseuodisk and mid-plane material. These characteristics naturally lead to the ``Hubble''-law behavior observed at large distances \citep{shang2020}.

Two prominent features can be seen in these outflows: an axial jet from the primary wind and a shell mainly from the swept-up ambient material. 
For $n\lesssim 2$, the shell can be quite collimated, producing jet-like molecular outflows. 
Alternatively, the shell has a large opening angle as in the ``classical'' molecular outflows, depending mainly on the angular distribution of the initial ambient material density 
as in the $n\gtrsim 4$ cases. These two observed classes of molecular outflow traditionally required separate and incompatible interpretation using either a jet- or wind-driven model. They can, however, be unified using this theoretical framework.

\subsection{Observed Features and the Unified Wind Model}
\label{subsec:unified_wind_model}

In this work, four representative ALMASOP sources are compared with the unified wind model. The morphological and kinematic structures within these outflows resemble those predicted by the model framework and numerical simulations outlined by \citet{shang2020, shang_PII}. The characteristic features predicted are diagnostic of the leading physics generic to such systems. However, the exact line intensities will depend on stellar parameters, environment, age, inclinations, and detailed time evolution.

The unified model establishes the required physics and structures within an elongated magnetized wind-blown bubble as a magnetized wind interacts with its magnetized surrounding material or envelope, collapsing or not \citep{shang2020, shang_PII, shang_PIII}. As required by the hydrodynamics of bubble theory advanced by \citet{KM1992A,KM1992B}, several layers of structures form within a bubble, from the innermost primary wind cavity to the outer ambient envelope: the reverse shock, the tangential discontinuity, and the forward shock. The reverse shock marks the interface between the primary and the compressed (shocked) wind material. The forward shock marks the interface between the compressed and unperturbed ambient material. The tangential discontinuity separates the compressed wind and the surrounding material and is subject to substantial shear forces and instabilities, resulting in filamentary fingers and vortices along the interface and penetrating deeply into the compressed wind and ambient media \citep{shang2020}. Additionally, the magnetic force generates pseudopulses in the compressed regions. This amplifies the fingers and vortices to coalesce into larger filamentary or shell-like structures \citep{shang2020,shang_PII}. Some of these large pseudopulses can converge toward the jet axes and alter the post-shock jet density and velocity, creating the impression of apparent mild pulses that fluctuate around the projected jet peak blobs. The innermost reverse-shock cavity, the multiple filamentary shells, and the outermost compressed ambient medium constitute the overall impression of the nested shell structure of the magnetic bubble, which can take on a range of shapes from spindles to cones.

The potential outflow structures can be characterized and parametrized by the properties of the wind and the surrounding envelope. The primary wind is characterized by the toroidal field strength in the wind in terms of Alfv\'enic Mach number $M_\mathrm{A}$. The ambient envelope structure is characterized by the flattening and the poloidal field strength threading the singular isothermal toroid, labeled by $n$ (see discussion in last sub-section \ref{subsec:history}, \citealt{li1996b}, and \citealt{shang2006}). The toroids evolve by increasing their deviation from the spherical singular isothermal sphere \citep{shu1977}, forming pseudodisks by flattening from the inner portions \citep[as shown by][]{galli93a,allen2003a,vaisala2023}. \citet{allen2003a} demonstrated how toroids open up as they evolve, determining the length-to-width ratios and opening angles of the resulting outflows. The relative sizes of the reverse-shock cavities and the magnetic interplay structures are related to the strengths of the toroidal field in $2<M_\mathrm{A}< \infty$ \citep[e.g.,][]{shang_PIII}. For a typical outflow from a low-mass system, $6\lesssim M_\mathrm{A}\lesssim 90$.

\begin{figure*}
\plotone{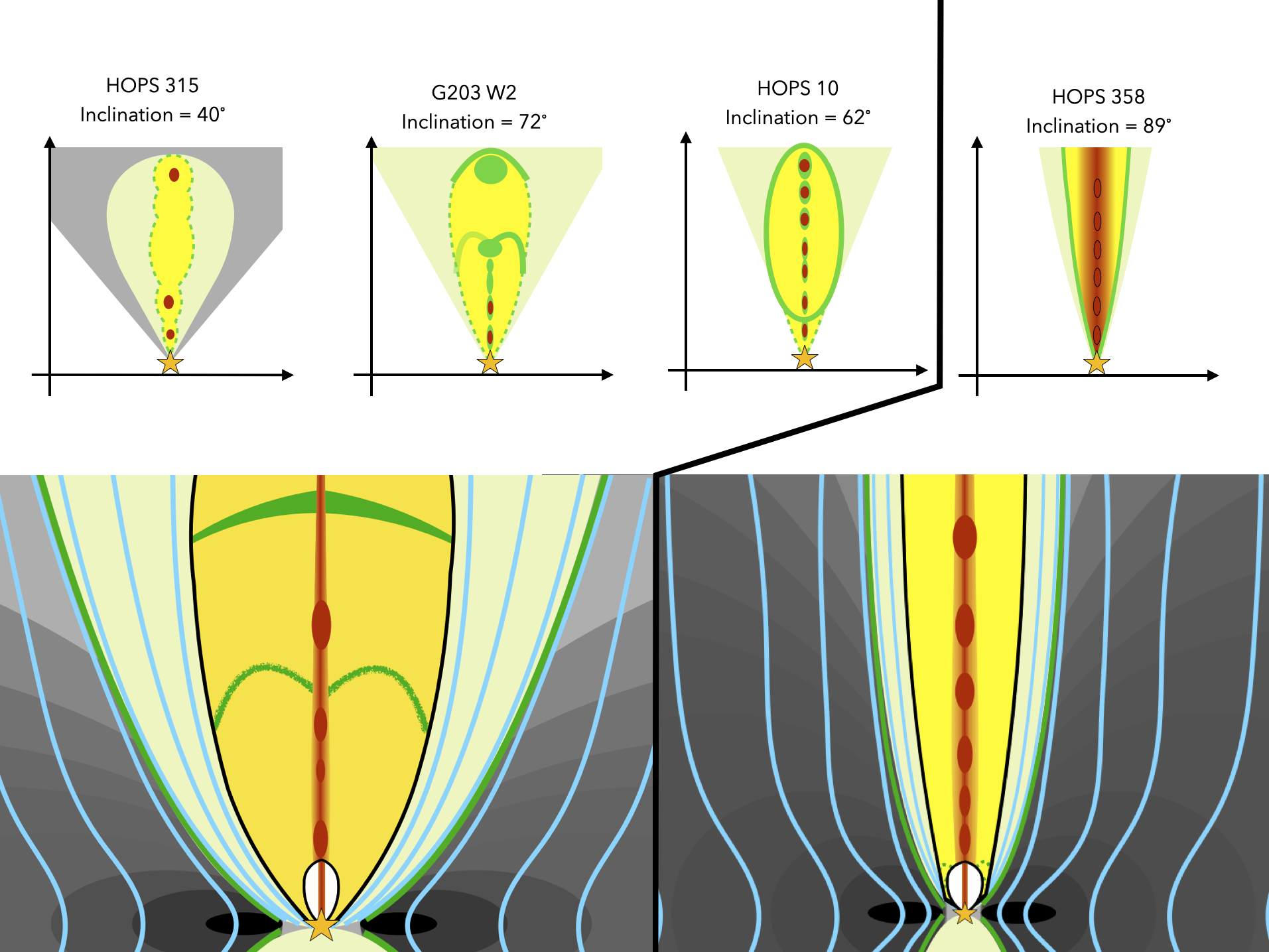}
\caption{Top panels: Schematics of observed components from the four representative ALMASOP sources. The three panels on the left illustrate cases with larger opening angles, indicating a more evolved status, and the right one presents a case with a smaller opening angle, indicating a younger, less evolved status.
Bottom panels: The schematic views of the general structures of the ALMASOP outflow systems under investigation, as described in Section \ref{subsec:unified_wind_model}.
The colors in all the panels encode information about the outflow regions following the descriptions by \citet{shang_PII,ai2024} as follows: compressed wind region (yellow), compressed ambient region (light green), and ambient material (gray). The mixing structures and SiO knots along the jet axis are marked in green and red within the compressed wind region, bounded by a black line, imitating the tangential discontinuity.
In the bottom panels, the free wind region and the reverse shock cavity are small-sized but important features at the base and marked in white. The magnetic field within the compressed and ambient regions is shown with blue lines.}
\label{ALMASOP_Outflows_Schematics}
\end{figure*}

The schematics in Figure \ref{ALMASOP_Outflows_Schematics} demonstrate the features seen in the unified model proposed by \citet{shang2020} for a jet-wind-outflow system emerging from a surrounding toroid and also revealed by the representative sources of this work. The flattened ambient structures imitate the toroids collapsing toward the inner denser part of the envelope. The four sources are sorted by the opening of their base cavities, characterized by the gradual opening of the toroids. The sources HOPS 315 and G203W2 appear to have larger base opening angles in the low-velocity CO components and wider compressed ambient regions in comparison with HOPS 10 and HOPS 358, as shown schematically in the top panels.

Along the outflow structures of the compressed wind, a large enhancement in density and emission occurs close to the axis, which naturally produces a jet-like feature. More localized enhancements along the jet take the form of knots or blobs, sometimes appearing near the end of arch-shaped filaments. These are a manifestation of the pseudo-pulsed structure, connecting the axial regions and the outer boundary of the compressed wind region, which is separated from the compressed ambient region by a theoretical tangential discontinuity. This boundary also coincides with the separation between the toroidally-dominated wind and poloidally-dominated ambient fields. 

\subsubsection{Unified Wind Model and ALMASOP Sources}\label{subsubsec:unified_wind_applications}
Here, we briefly discuss the connection between the physical features found in the unified wind model and the observations of the four representative ALMASOP sources, as shown in the top panels of Figure \ref{ALMASOP_Outflows_Schematics}:

\paragraph{HOPS 315} The CO emission shows an inner part tracing the compressed wind material and the interface between the wind and the environment. 
The arc-like structures produced by interaction are mainly seen in CO\@. The SiO emission mainly traces the axial region as part of the stratified wind density within the compressed wind. The most distant CO shells represent the outer boundary of the compressed ambient material. The number of observed knots is inconsistent with the number of large bow shocks (e.g., 5 vs.\ 2 in the red-shifted and 4 vs.\ 2 in the blue-shifted lobe), potentially requiring an additional non-bow shock origin mechanism.

\paragraph{G203W2} The SiO knots are brighter in the lower portion along the jet axis where CO also appears. On the other hand, the SiO emission is weak downstream, where CO emission is brightest. If all the knotty structures had been formed through jet bow shocks, the SiO would be expected to overlap with the CO knotty structures. Moreover, the CO emission shows an asymmetric axial distribution most likely attributed to ambient material mixed into the outflow lobe through entrained and compressed ambient material.

\paragraph{HOPS 10} A series of SiO knots and CO blobs are seen in the fastest velocity channel, but no trace of these bow-shock shells are observed in CO\@. In this source, the lower velocities exhibit bubble-shaped structures. These suggest an origin with dual characters of jet and wide-angle wind and support structures within a magnetized bubble arising from magnetic interplay with the environment.

\paragraph{HOPS 358} While a number of SiO knots can be seen, no episodic shells can be identified in CO\@. The SiO emission becomes less knotty at the systemic velocity and appears to fill the CO conic structure. These features are reminiscent of structures within the jet-containing X-wind, where bubble structures form due to magnetic interplay with the surroundings.

\subsubsection{Reverse Shock Cavities} \label{subsec:RSCavity}

Identifying the innermost cavity formed by the reverse shock within the outflow lobe, and predicted by the bubble structures in the \citet{shang2020} unified model, is one of the most significant findings in this work. This feature is most easily identified through the parallel PVDs when the velocity structures can be resolved \citep{shang_PII}.

The properties of an outflow change during the crossing of the reverse shock. Densities increase while velocities decrease and can be deflected and converged. A correlation of these effects is generally seen in parallel PVDs as a roughly triangular region at low position $|z|$ near the base, marked by yellow dashed lines in Figures \ref{HOPS315_PVDs}, \ref{HOPS10_PVDs}, \ref{G203W2_PVDs}, and \ref{HOPS358_PVDs}.
Voids or faint emission features generally appear on the low $|z|$ side of each trace, while much stronger emission is regularly seen on the high $|z|$ side. The jet is often an exception to this pattern, clearly visible for lower $z$ values in HOPS 358 and the blue-shifted side of both HOPS 315 and G203W2. 
These are the cases within this representative sample where the pristine SiO jet is potentially being observed.

This triangular delineation in the parallel PVDs, extending from low to high velocities, can identify the location and presence of the reverse shock as a place where the flow changes from a simple free wind into an interacting outflow. The wide-angle wind material within this region may be too deeply embedded or not dense enough to be detected in HOPS 10 and the redshifted sides of HOPS 315 and G203W2. However, in all cases SiO emission is apparent beyond the yellow dashed regions. Hence, the reverse shock is shown to help converge, confine, and enhance the density of the central jet component of the wide-angle wind to meet the necessary physical requirements for SiO excitation.

In all four sources, the most salient jet features observed correspond to the shocked jet located after the crossing of the reverse shock. Given that the wind and jet patterns observed downstream of the reverse shock have been processed, they should not, therefore, be interpreted directly as the pristine jets or winds but rather as shocked flow structures (see Section \ref{subsec:conundrums_conventional} for additional details).

\subsubsection{Nested Kinematic Structures}
\label{subsec:pv_discussion}

The structures related to the observed kinematic features in the representative ALMASOP sample have been demonstrated by \citet{shang_PII} through parallel and transverse PVDs. Filamentary structures tracing the magnetic interplay stretch from the cavity to the jet component (as indicated by the white dotted lines in the upper-left panel of Figure \ref{HOPS315_PVDs}). 
Some large fingers penetrate the axial jet region and enhance the material there, mimicking density and velocity variations and creating large knot-like wide CO and SiO blobs. The magnetized compressed ambient region is traced by the distribution of material with very low velocity along the $v=0$ position axes. The compressed ambient material and the unseen poloidal field constitute the spatially extended LVC as the outermost part of the nested shell structures. 

\begin{figure*}
\plotone{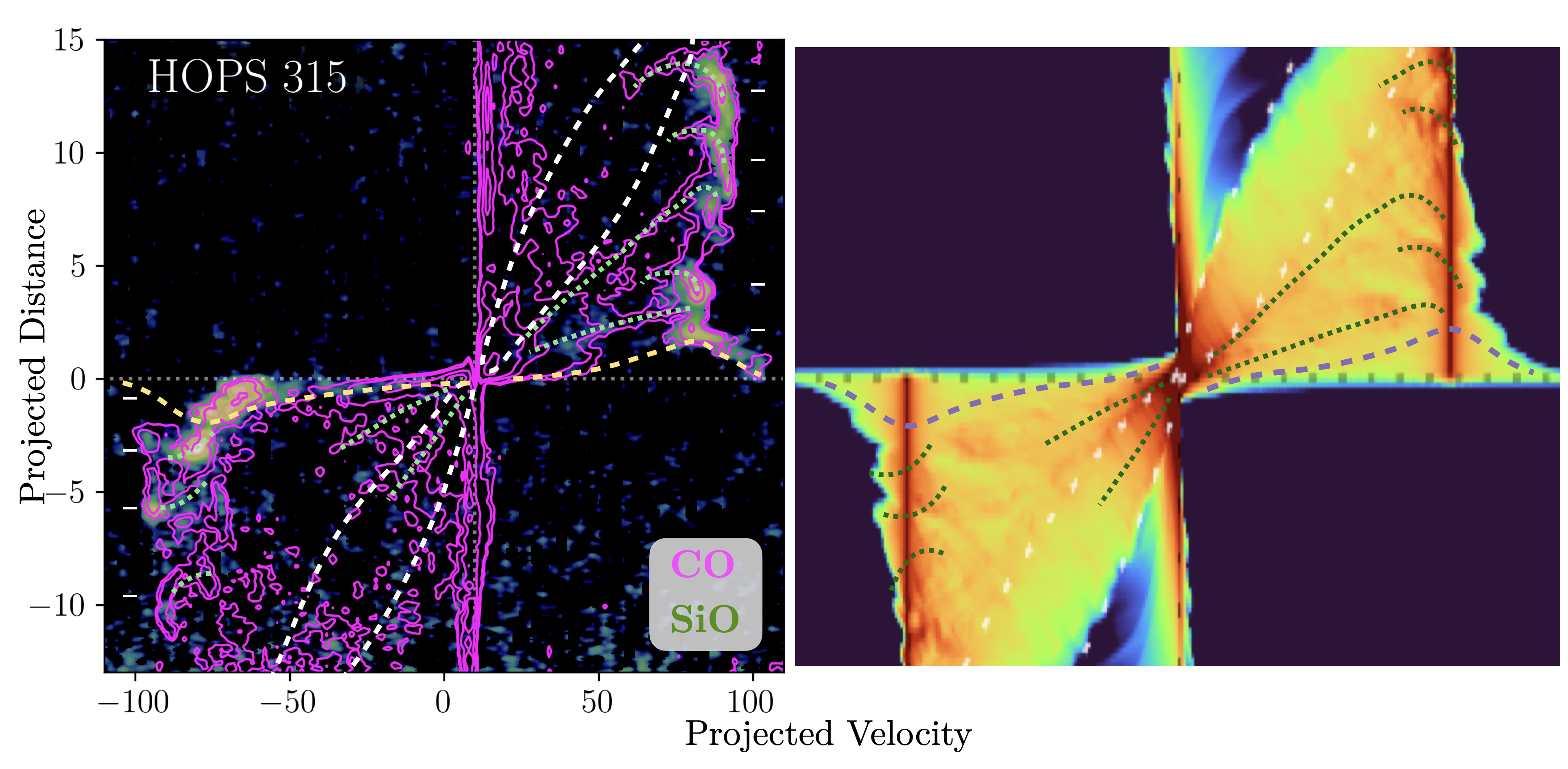}
\caption{An example of comparison for the parallel PVD of combined CO and SiO emissions of HOPS 315 (left panel) with synthetic column density in PV space (right panel) as shown in \citet{shang_PII}. In both panels, yellow dashed lines near the jet base delineate the reverse-shock cavities, the filamentary structure is traced by dotted lines connecting the EHV knotty structure toward the LV region, and the white dashed spindles or ellipses connecting the origin to the tip of the jet indicate the theoretical trace of the momentum-conserving shell in the outflow system. Jet features are very clear in both observation and simulation as EHV structures. The vertically extended, very low-velocity features along the position axis are produced by the compressed ambient material when the ambient medium is magnetized.}
\label{ParaPVD_SynthComparison}
\end{figure*}

As an example, Figure \ref{ParaPVD_SynthComparison} shows the juxtaposition of HOPS 315 ($i=40\arcdeg$) and an example synthetic PVD produced by \citet[][case using $M_\mathrm{A}=30$ and $n=4$ toroid for an initial wind velocity of $100\kms$ and $i=45\arcdeg$]{shang_PII}. The observed kinematic structures revealed by the parallel PVDs can be identified and accommodated with an approximate bracket of parameter and inclination angles between $30$--$60\arcdeg$, and a mass-loss rate of $10^{-6}{\mathrm M}_\odot/$yr. These inclinations are required such that the jet and wide-angle wind velocities can be properly separated from the very low-velocity compressed ambient material \citep{shang_PII}. 
Similar parallel PVDs are observed for HOPS 10 and G203W2, both of which lie within a similar range of inclination angles $\sim 45^\circ$ and qualitatively correspond to similar PVDs produced by the unified model simulations \citep[][Section 7]{shang_PII}. However, the fits prefer $1\lesssim n\lesssim 4$ and $6\lesssim M_\mathrm{A}\lesssim 30$. 

In the parallel PVDs, the spatially extended LV, IV, and EHV components are present in CO.
These PVDs exhibit the nature of a highly magnetized wind interacting with ambient medium, forming a thick compressed wind region as extended CO emission distributed between the LV and EHV as a ``rectangle'' shaped region on each side (blueshifted and redshifted). This feature results from magnetic interactions that disrupt the conventional momentum-conserving approximation of thin shells.

The conventional cavity wall is given by the theoretical ``spindle-like'' curves in the hydrodynamical momentum-conserving thin shell and traced by the dotted spindle-shaped curves connecting the origin to the jet's tip in Figure \ref{ParaPVD_SynthComparison}. The conventional hydrodynamical momentum-conserving curves do not fit the real features of these parallel PVDs, which is to be expected based on the magnetized version of the bubble structures predicted by \citet{shang_PII}. 

The transverse PVDs, when combining the view of the blueshifted and redshifted lobes, provide a glance at the kinematic information across the lobes. As in the parallel PVDs, the regions close to the systemic velocity are dominant in ambient material and therefore bright in CO emission, on top of the self-absorption features at the systemic velocity \citep[see, e.g.,][]{dunham2014b}. 

Next to the contribution of the ambient material, the most evident structure is the rhombus-like region filled in by CO emission. 
The SiO jets appear as localized blobs concentrating at the average jet velocities on the jet axis in the shocked compressed region; however, their velocity is somewhat reduced from the deprojected velocity of a full pristine jet.

Within the region, the CO emission is confined but filled in an oval representing the compressed (shocked) wind material after passing the reverse shock, denoted by the fastest CO emission on the edges of the oval's major axis. The entire CO emission may look more like a concave rhombus than a filled oval due to the mixing of multiple velocity components encompassing the compressed wind and ambient material within the outflow lobes \citep[as shown in Section 8 of][]{shang_PII}. The fact that the CO fills in a clearly defined region, oval or rhombus, strongly suggests a connection to entrained material from the ambient envelope.

\subsubsection{Origins of the SiO Emission} \label{subsec:sio_formation}

In these representative sources, the CO and SiO do not overlap significantly, which suggests that the origins of the SiO and CO emission may differ. The SiO potentially traces material from the pristine wind, advected across the reverse shock into the compressed wind region. This idea is supported by the observation that most SiO emission becomes visible after the location associated with crossing the reverse shock cavity. It is an inner part of the compressed wind region as a continuation of the underlying high-density innermost jet region at the center of the wide-angle wind. On the other hand, the CO is not expected to originate in the pristine wind entirely. The stronger CO emission appears to encompass the SiO emission, with an upper velocity capped by the SiO downstream of the reverse shock in the region associated with the compressed wind and distributed from moderately high velocities down to the ambient values. This salient correlation indicates that CO is tracing the magnetized interplay with the wider portions of the compressed wind region and, through that, also with the regions of compressed and uncompressed ambient media and the regions of mixing with entrained ambient CO.

Observations of SiO emission from Class 0/I sources suggest an enhancement of SiO molecules in the gas phase by several orders of magnitude \citep{tafalla2011}. This gas-phase SiO abundance enhancement is often attributed to two main mechanisms, one associated with the SiO being in the gas phase as part of a high-density molecular primary wind \citep[e.g.,][]{glassgold1991} launched from within the dust sublimation radius, and the other one associated with SiO being sputtered from dust grains destroyed via shocks in a rather dusty supersonic wind \citep[e.g.,][]{schilke1997}.

In the representative sources investigated here, the SiO emission is found to emerge beyond the first cavity, which is interpreted as the reverse shock cavity of the magnetized wind bubble. Either of the above scenarios can be applied to the emergence of this SiO emission. 
The initial molecular hydrogen number density might be below the critical density required for SiO transitions due to relatively low mass-loss rates in these sources. After the reverse shock, the postshock compression might achieve the critical density. Alternatively, if the wind is launched with SiO still depleted onto dust grains, mild to moderate shock velocities \citep[$\gtrsim10\kms$,][]{schilke1997} may be able to destroy the grain mantle and release the SiO into the gas phase.

\subsubsection{Apparent and Real Episodicity}
\label{subsec:episodicity}

Multiple knots and shells are present in many protostellar outflow sources. The origin of repetitive, sometimes quasi-periodic, structure can be explained in at least two ways.
First, the quasi-periodicity may be due to real episodicity of the jet and wind, a variability in the accretion processes through the disk, linking the infall variability directly to the outflow formation process. Second, the outflow itself may introduce instabilities that lead to quasi-periodic structures. \citet{shang2020} studied magnetized outflows without explicit wind/jet variability and found the emergence of apparent episodic structures as a result of wind-ambient magnetized interplay \citep[see, for example, the kinematic signatures shown by][]{shang_PII}. These interplay structures are called ``pseudopulses'', and can happen both in time-dependent and time-independent sources.

Each of these two causes of episodicity has a strong physical basis, which leads to consequences that can be discriminated through observation. Within our sample of four objects observed in CO and SiO, we suggest that both explanations are necessary, as we identify clear cases of pseudopulses and potentially one clear case of imposed real episodicity.

Regarding pseudopulses, large-scale filaments propagating from the outer shell inward toward the axis can be produced each flow dynamical time. 
In contrast, small-scale density fluctuations can occur with quasi-periods shorter by a factor of a few compared to these large-scale filaments. An almost constant high-velocity velocity centroid dominated by brightness is a characteristic feature of the magnetic pseudopulse in an idealized axisymmetric system. Lower-velocity light filamentary threads can be seen to link structures of different velocities to the lower-velocity ends of the blobs, as shown on the synthetic parallel PVDs in Figure \ref{ParaPVD_SynthComparison} and in general, Figure 16 of \citet{shang_PII}. 

\paragraph{Representative Sources with clear pseudopulses}

For HOPS 10, a series of SiO knots extending up to $\sim15\arcsec$ can be identified along the jet axis.
Despite internal structures in the PV space, the velocity centroids of the SiO knots essentially follow the jet velocity of $v_\mathrm{LSR}$ of $\sim 65\kms$, as shown in the parallel PVDs in Figure \ref{HOPS10_PVDs}. Except for a decrease to $\sim50\kms$ at $\sim5\arcsec$, the velocity centroids do not vary beyond $\pm 6\kms$, or $\pm10\kms$ after deprojection.
Similar numbers can be presented for the redshifted side of HOPS 315, which is compared to a pseudopulse model in Figure \ref{ParaPVD_SynthComparison}. There, we can see that the pseudopulse mechanism can produce a variety of knots, all virtually at the same PV velocity as the jet with small oscillation amplitudes in PV space, with the filamentary connections to the lower velocity components. These are signatures in PV space of the formation mechanism of pseudopulses \citep{shang_PII}. The filamentary characteristics of the pseudopulse model are present in the parallel PV of all four sources, not excluding G203W2. 
The constant velocity of the knots is combined with a steady jet velocity feature in HOPS 10 and HOPS 315 (redshifted sides) and in HOPS 358 (both sides). HOPS 10 (blueshifted side) is too weak for clarity, and HOPS 315 (blueshifted side) has an enticingly complex SiO structure, perhaps due to SiO excitation along the reverse shock cavity or magnetic acceleration as in \citet{shang_PIII}, challenging both the pseudopulse and the traditional episodic approach.

Head--tail kinematics associated with real episodicity are observed and can be modeled by strong velocity or mass-loss variations, as demonstrated by \cite{wang2019} for the large, intriguing patterns found in IRAS 04166+2706. Within each of these formed (large) knots, there is an apparent negative velocity gradient downstream, e.g., each knot being slower at larger distances from the star (``head'') than closer to the star (``tail''). These velocity patterns have also been seen in the Class 0 EHV jets from both IRAS 04166$+$2706 \citep{santiago-garcia2009,tafalla2017,wang2019} and L1448C-N \citep{hirano2010,yoshida2021,toledano-juarez2023}.

Such episodic ejections have been conventionally invoked to explain the creation of observed internal working surfaces or bow shocks, in which the faster ejecta catch up with previously ejected and existing slower ejecta, \citep[see, e.g.,][]{santiago-garcia2009,wang2014,jhan2016}. This scenario has also been applied to several EHV jets within the ALMASOP sample to determine the ejection history and inclinations of the outflows \citep{jhan2022}. 
Possible mass accretion episodicity of the stellar system \citep[e.g.,][]{park2002,contreras-pena2020,lee2021_Transient} may contribute to the generation of pulses and the formation of the knotty structure in the jet \citep[see analysis by, e.g.,][]{dutta2024}. Both pseudopulses and real episodicity may be present in the same astronomical object. In that case, the appearance of episodic pulses can be enhanced by the magnetic interplay that induces the pseudopulses or vice versa, with the overlap of each other signatures in velocity and position spaces of the various features of the object.

\paragraph{G203W2 and its episodicity}

For G203W2, the velocity structure may be grouped into three regions along the flow. At $z<5\arcsec$, the velocity dips of the SiO knots in the compressed wind/shocked jet region can be traced. Unlike fluctuating velocity around the jet centroids, as seen from the other three sources with clear pseudo-pulses, these consecutive velocity drops in a series of knots as a trend suggests velocity variations in overall flows. The knotty CO and SiO at $\sim8\arcsec$ are linked to an off-axial CO arc, with an apparent velocity jump relative to the upstream material, which appears like the wiggling of the jet, which interacts with the adjacent material. 
However, the jets do not follow the wiggling pattern in all the channels. On the other hand, envelope entrainment can bring surrounding material into the outflow lobe, and the non-axisymmetric behavior can result from nonisotropic material distribution or inhomogeneous interaction with the ambient medium. 
At $\gtrsim 12\arcsec$, a large arc-like structure is associated with a velocity centroid decrease in CO and faint SiO PVD. This structure may be tracing an internal bow shock along the jet axis. 
Overall, this source's velocity pattern exhibits systematic variations that suggest real episodic ejection may be at play.
\smallskip

In summary, we interpret apparent and real pulses as complementary rather than opposing each other, and a concrete example of this can be found in G203W2, as demonstrated above. The filamentary structures in PV, which the pseudopulse model explains so well, seem universal in our sample, while real episodicity seems necessary in some sources. 

\subsection{Observed Features Confronting Unified and Conventional Models} \label{subsec:conundrums_conventional}

We have combined the signatures of CO and SiO emission to elucidate morphological and kinematic structures within the representative ALMASOP outflow sample. These features are identified through integrated maps, channel maps, and PV diagrams parallel and transverse to the outflow axis.

At first glance, the integrated maps shown in Figure \ref{fig:jet-cavity_maps} appear to reveal a conventional jet-shell morphology for all four sources. The morphological coexistence of LV cavities and EHV jets in CO emission and the dominant role of EHV jets in SiO emission can be seen. The CO channel maps, however, show distinct morphological characteristics in each of the channels. In HOPS 315 (Figure \ref{HOPS315_ChMap}), for example, from fastest to slowest, the four representative channels show jet-like, bow-shaped, shell-like, and a cavity with a bubble-shaped shell structure. A straightforward explanation requires assuming four different types of driving, one for each channel. The fastest channels may be attempted with a jet bow-shock scenario, the middle channels may be presented with a wind-driven shell structure, and the slowest channels may be a mixture of slow, wide-angle (disk) wind and a bubble-like shell.

The other three sources show a similar mixture of morphological features in different velocity channels. Conventional explanations based on various types of driving could be constructed, one per feature and velocity channel, but missing the opportunity to present a unified cause. For G203W2 (Figure \ref{G203W2_ChMap}), there are cone-like limb-brightened shells in LVC and bubble-like structures with knotty structures at higher velocities. For HOPS 10 (Figure \ref{HOPS10_ChMap}), the brightest CO emission appears as a cone-shaped shell in the lower velocity channels, with a diffuse emission outside of the shell in the lowest velocity channel and a more jet-like structure in the high-velocity channel. This nested structure appears similar to that of DG Tau B \citep{devalon2020}.

HOPS 358 (Figure \ref{HOPS358_ChMap}) also exhibits a nested structure. The SiO and CO emission appear to fill within a narrow limb-brightened conic CO structure. Outside the conic emission lies a diffuse emission region with a larger opening angle. The narrow structure is nested within the wider-opening structure due to the edge-on orientation in the lowest-velocity channels. Being edge-on, projection effects do not allow such a clean separation of structures according to the channel maps.

In the parallel PVDs for all sources, several kinematic features can be seen to coexist. For the EHV region, a series of knotty structures can be seen in both SiO and CO emission, which may be accounted for by velocity variations from a narrow jet \citep{jhan2022} or an axially concentrated wide-angle wind \citep{wang2019}. In the LV region, the ambient material dominates the CO emission, extending across the field of view. The velocity region involves complex structures that have no correspondence with the conventional jet or wind models. The area between the LV and EHV components is filled with diffuse CO emission, occasionally showing filamentary structures connecting the two components. No clear identifications of possible fitting of with a potential  
parabola-like structure \citep{lee2000} can be made. Moreover, the triangular cavity shown on the parallel PV space (e.g., Figures \ref{HOPS315_PVDs} and \ref{G203W2_PVDs}) lacks explanations in either of the conventional pure hydrodynamical jet-driven
or wind-driven scenarios. The range of the velocity convergence in the EHV region is also incompatible with the slow molecular (``disk'') wind scenario. (see below the paragraph on the difficulties with this scenario).

The observed reverse shock marks the region in which the wind starts its interactions. Crossing this shock will slow down the wind velocity, providing a natural explanation for the slower velocity components dropped from the pristine full jet/wide-angle wind velocities. 
The ambient material velocity may increase in magnetized interactions within the compressed materials, allowing material mixing and a variety of phenomena unaccounted for in conventional scenarios.

A commonly adopted alternate and popular explanation for observed slowly moving features is the invocation of a very slow molecular disk wind. Examples of this approach \citep[e.g.,][]{louvet2018,tabone2020,devalon2020,devalon2022,lopez-vazquez2024} are also able to fit the low velocities of the LVC\@. Alternatively, the unified model produces the low velocity self-consistently through naturally occurring reverse shock fronts. Indeed, even for a hypothetical slow disk wind, there is the expectation of interaction (shocks) within the outflow; thus, one must be careful in interpreting slow-moving gas to understand where and how it may possibly arise. The unified model shows that physical processes within outflows should interact, and the ALMASOP sample indicates this interaction indeed takes place at all scales, close to and further away from the source.

The unified model well explains the rich and complex kinematic structures in the four representative samples. The intertwined features of jets, winds, and outflows within an environment suggest that all outflow model scenarios require a unified coexistence. This is currently best exemplified in the literature by the magnetized interplay of \citet{shang2020, shang_PII}.

\subsection{Disentangling Features in the Larger ALMASOP Outflow Ensemble} \label{subsec:ensemble}

The four selected representative outflow sources stand out from the larger ALMASOP sample in that they possess a full set of distinct kinematic features anticipated within an outflow system. The overall features, including the jet-shell morphology, wide-angle wind nature, and the compressed wind and ambient material residing within the intermediate velocity range, manifest themselves via the channel maps and PVDs. The CO and SiO emission diagnostics can also clearly identify and discriminate those features through density and excitation selections.

These sources may also provide a generic investigation into the kinematic and morphological structures revealed by the combined analysis of CO and SiO emission since they cover various velocity components at a range of observed inclinations. The analysis can also be applied to the remaining 15 ALMASOP sources showing both CO and SiO emission despite various morphologies and velocity coverage. 

Excitation conditions also vary among different outflow sources. Within the overall 31 outflow sources in the ALMASOP sample, the SiO emission is undetected toward 12 sources. As discussed in the previous subsection, the SiO emission is sensitive to the density and excitation conditions in the primary winds. Despite the different physical conditions among the sources, the analysis here can still fit into the general analysis and interpretation framework. 

We also note that the data are limited in sensitivity due to the relatively large distance to the sources. As more detailed structures are revealed, observations with higher sensitivities are required to further determine the nature of the fine structures and further compare and analyze them against the proposed model framework.

\section{Summary} \label{sec:summary}

We selected and highlighted four out of the 19 outflow sources in the ALMASOP sample with both $^{12}$CO ($J=2-1$) and SiO ($J=5-4$) emission associated with emerging outflows, namely HOPS 10, 315, 358, and G203W2. The circumstellar environment has also been investigated through N$_2$D$^+$, C$^{18}$O, and H$_2$CO\@. Kinematics and morphological structures are identified by examining the channel maps and position--velocity diagrams in the parallel and transverse directions of the outflow axes in the context of the unified model framework based on the respective predictions in physical mechanisms, morphological and kinematic features \citep{shang2020,shang_PII}.

We identify the EHV jet, LV cavity walls, and nested structures within the cavity. The jet features appear in the highest velocity channels in the datacubes as a collimated jet component. The lower-velocity gas component around the systemic velocities exhibits shell-like morphology surrounding the jet. Within this subset of four outflows showing SiO jets and CO outflows, clear jet-bearing wide-angle wind features can be identified through the large line widths of SiO emissions close to the driving sources. 

New features are also identified beyond the classical features of jet--shell structures in this outflow sample. Filamentary structures are found in the intermediate velocity region, connecting the low-velocity emission and the knotty jet-like structures. These features connect the outer cavities with the inner axial jets. At the bases, large line widths are associated with an innermost cavity of similar velocity dispersion, and kinematic features correlate with the nested morphology from the inside out. The newly revealed features can be accommodated within the theoretical framework predicted by the unified magnetized bubble. The observed outflow structure is a bubble structure formed through the interaction between a magnetized wide-angle wind and the ambient environment.

The reverse shock and its cavity have been identified in all four sources (Section \ref{subsec:RSCavity}). The region immediately downstream of this shock is the compressed wind region, where most of the newly identified phenomena occur. Most of the SiO jet-like emissions in this work arise in the compressed wind region , which appear on top of the approximate triangular-shaped region,  where the flow properties change across the reverse shock due to enhanced excitation conditions,
or to sputtering from shock-disrupted dust grains (Section \ref{subsec:sio_formation}). 
Most visible SiO jet features correspond to this shocked jet portion, including the knotty appearances, pseudopulses, multi-shell filamentary structures, and possible episodicity. 
Hence, the unified model offers a new framework for understanding and interpreting jet emissions. 

The kinematic structure of these four exemplary sources can be accommodated by the unified wind model for outflows in the framework of \citet{shang2020,shang_PII}. Such analysis can be extended to the other 15 ALMASOP outflows with CO and SiO emissions and 12 sources with only CO outflows for the differences in excitation conditions and evolutionary status. The general structure of the ALMASOP outflow sources provides an opportunity to examine the systematic properties of observational features obtained by ALMA capabilities and their subsequent confrontations with numerical modeling.

%\begin{acknowledgments}

\vspace*{1cm}
\noindent

The authors acknowledge grant support for the CHARMS group from the Institute of Astronomy and Astrophysics, Academia Sinica (ASIAA), and the National Science and Technology Council (NSTC) in Taiwan through grants 111-2112-M-001-074-, 
112-2124-M-001-014-, 112-2112-M-001-030-, and 113-2112-M-001-008-. The authors acknowledge the access to high-performance facilities in ASIAA\@. 
D.J.\ is supported by NRC Canada and by an NSERC Discovery Grant.
M.G.R. is supported by NOIRLab, which is managed by the Association of Universities for Research in Astronomy (AURA) under a cooperative agreement with the National Science Foundation.
P.S. was partially supported by a Grant-in-Aid for Scientific Research (KAKENHI Number JP22H01271 and JP23H01221) of JSPS. This paper makes use of the following ALMA data: ADS/JAO.ALMA\#2018.1.00302.S. ALMA is a partnership of ESO (representing its member states), NSF (USA), and NINS (Japan), together with NRC (Canada), NSTC, and ASIAA (Taiwan), as well as KASI (Republic of Korea), in cooperation with the Republic of Chile. The Joint ALMA Observatory is operated by ESO, AUI/NRAO, and NAOJ.

%\end{acknowledgments}

\vspace{5mm}
\facilities{ALMA}

\software{astropy \citep{astropy2013,astropy2018,astropy2022},  
          Numpy \citep{Numpy},
          Scipy \citep{2020SciPy-NMeth}, 
          OpenCV \citep{opencv_library},  
          SpectralCube \citep{Spectral_cube},
          Matplotlib \citep{hunter2007}
          }

\bibliography{ALMASOP}{}
\bibliographystyle{aasjournal}

\end{document}